\documentclass[twocolumn,aps,pr,superscriptaddress,preprintnumbers,nofootinbib,10pt]{revtex4-2}
\usepackage{multirow}
\usepackage{amsmath}
\usepackage{amssymb}
\usepackage[dvipdf,dvips]{graphicx}
\usepackage{color}
\usepackage{hyperref}
\usepackage{url}
\usepackage{slashed}
\usepackage{subfigure}
\usepackage[usenames,dvipsnames]{xcolor}
\usepackage{amsmath}
\usepackage{amsfonts}
\usepackage{float} 
\usepackage{amssymb}
\usepackage{graphicx}
\usepackage{array}
\usepackage{epsfig}
\usepackage{graphics}
\usepackage{euscript}
\usepackage{slashed}
\usepackage{epstopdf}
\usepackage[utf8]{inputenc}
\allowdisplaybreaks
\usepackage[normalem]{ulem}
\usepackage{pifont}
\usepackage{dsfont}
\usepackage{MnSymbol}
\usepackage{verbatim}
\usepackage{graphicx}
\usepackage{latexsym}
\usepackage{tikz-feynman}
\usepackage{cancel}

\begin{document}
	
\title{Yang-Mills theories with scalar matter: a first foray using the Curci-Ferrari model}

    \author{Gustavo L. L. W. Levy} \affiliation{CBPF $-$ Centro Brasileiro de Pesquisas Físicas, Rua Dr. Xavier Sigaud 150, 22290-180, Rio de Janeiro, Brazil}
    \affiliation{Centre de Physique Théorique, CNRS, Ecole polytechnique, IP Paris, F-91128 Palaiseau, France}
    
	\author{Philipe De Fabritiis} \affiliation{CBPF $-$ Centro Brasileiro de Pesquisas Físicas, Rua Dr. Xavier Sigaud 150, 22290-180, Rio de Janeiro, Brazil}
    \affiliation{KU Leuven Campus Kortrijk-Kulak, Department of Physics,	Etienne Sabbelaan 53 bus 7657, 8500 Kortrijk, Belgium}

	\author{Marcela Peláez} \affiliation{Instituto de Física, Facultad de Ingeniería, Universidad de la República, J. H. y Reissig 565, 11000 Montevideo, Uruguay}

    \author{Urko Reinosa} \affiliation{Centre de Physique Théorique, CNRS, Ecole polytechnique, IP Paris, F-91128 Palaiseau, France}
    

\begin{abstract}

The Curci-Ferrari (CF) model provides an efficient perturbative framework for describing several aspects of the infrared (IR) dynamics of Yang-Mills (YM) theories. We extend this framework to YM theories coupled with scalar matter and investigate its main properties, with particular emphasis on the one-loop two-point correlation functions of scalars, gluons, and ghosts. We analyze the renormalization group (RG) structure of the model, computing its beta functions and studying the corresponding RG flows for different scalar masses. We establish the existence of IR-safe trajectories along which perturbation theory remains under control and identify a novel fixed-point, while recovering the standard CF results in an appropriate limit. We then confront our results with lattice data through a simultaneous fit of the scalar, gluon, and ghost propagators, finding excellent agreement. Our results indicate that, within the approximations considered here, the CF framework gives a successful perturbative description of the IR regime of YM theories coupled with scalar matter.

\end{abstract}
		
\maketitle	

\section{Introduction}\label{SecIntro}

Yang-Mills (YM)~\cite{Yang54} theories stand as one of the most powerful achievements of theoretical physics, providing the mathematical backbone for the Standard Model of Particle Physics~\cite{tHooft07}. In particular, this framework encompasses Quantum Chromodynamics (QCD), the quantum field theory which successfully describes the strong interactions between quarks and gluons, the elementary particles which build all known hadrons~\cite{Gross23, Brambilla14}. In the ultraviolet (UV) domain, where asymptotic freedom~\cite{Gross73,Politzer73} ensures the reliability of perturbation theory, the success of QCD has been confirmed with remarkable precision in countless experiments. Yet, this triumph contrasts sharply with the problems posed by the infrared (IR) regime, where non-perturbative phenomena such as confinement remain without a complete analytical explanation despite the huge efforts of last decades~\cite{Greensite11}. Understanding the IR regime of YM theories remains a major challenge, that must be addressed to unravel the mechanisms underlying the dynamics of strong interactions.

An interesting way of probing the IR regime of YM theories is to study correlation functions for infrared momenta, a subject that has been intensively developed in the past decades, through many different approaches, such as the functional renormalization group and Dyson-Schwinger equations~\cite{vonSmekal97, Alkofer01, Aguilar08, Alkofer10, Huber20, Pawlowski04, Fischer09, Dupuis21, Cyrol16, Cyrol18}. In particular, a strongly reliable non-perturbative framework to access this regime is given by Lattice QCD~\cite{Wilson74, Creutz83}, which can provide first-principles numerical results for a great variety of quantities. In the past decades, lattice simulations revealed (at least) two striking features: i) the gluon propagator in Landau gauge saturates at a finite value, corresponding to an effective screening mass; ii) the gauge coupling, defined from the ghost-gluon vertex, remains finite and moderate at all scales~\cite{Bogolubsky09}. Needless to say, these astonishing features are in sharp contrast with the predictions of the standard Faddeev-Popov (FP) quantization~\cite{Faddeev67}. This discrepancy highlights the limitations of the conventional perturbative approach and underscores the need for alternative formulations to capture the non-perturbative features of the IR regime. 

The above-mentioned limitations of the FP approach can be traced back to the existence of multiple gauge configurations satisfying the same gauge-fixing condition, the so-called Gribov problem~\cite{Gribov78, Vandersickel12}. Indeed, the existence of such ambiguities implies that more than one representative per gauge orbit is being selected by the gauge-fixing condition, violating a key assumption of the method and undermining the standard FP quantization.

In a seminal work, Gribov devised a way to deal with these ambiguities, restricting the functional integral to a region free of (infinitesimal) Gribov copies, implementing such a restriction at first order through the famous no-pole condition~\cite{Gribov78}. A decade later, Zwanziger implemented the restriction at all orders in perturbation theory following another route, through the horizon condition~\cite{Zwanziger89}. Furthermore, it was proven that every orbit crosses this region~\cite{DellAntonio91} and that this problem is not a peculiarity of a particular gauge-fixing, but a more general phenomenon in YM theories~\cite{Singer78}. Interestingly enough, extending Gribov's no-pole condition to all orders, one recovers Zwanziger's horizon condition~\cite{Capri13}. 
Although improving the FP quantization, the Gribov-Zwanziger model was not able to reproduce the lattice results mentioned previously. The model was then improved to consider the emergence of certain condensates in the IR, leading to the so-called refined Gribov-Zwanziger (RGZ) model~\cite{Dudal08}. The RGZ action is local and renormalizable at all orders of perturbation theory, removes the (infinitesimal) Gribov copies from the functional integral and finds a good agreement with the lattice data for the gluon and
ghost propagator in the IR~\cite{Vandersickel11}. In recent years, there has been progress in the computation of correlations within the RGZ framework at one-loop order~\cite{Brito24, Brito25, Barrios24b, Brito23, Mintz18}, but this is still in its early stages.

Despite its appeal as a first-principles analytical approach for investigating the IR regime of YM theories, the RGZ framework has some important drawbacks. First, it only remove infinitesimal Gribov copies: the ambiguities related with finite gauge transformations are still there~\cite{vanBaal92}. Second, although the massive parameters present in the RGZ action can be fixed self-consistently through suitable gap equations, solving them explicitly to predict the behavior of the gluon propagator, for example, is a challenging task,\footnote{Not to mention that they require renormalization factors for composite operators which carry their own arbitrariness.} hence these parameters are usually treated as fitting parameters. Finally, performing loop computations within the RGZ action is cumbersome, as it introduces new mixings and vertices that cause a proliferation of Feynman diagrams~\cite{Brito24} and no true small expansion parameter associated to the condensate sector, thus restricting its efficiency. Therefore, it would be highly desirable to formulate an effective approach that captures the essential features of the IR regime while remaining efficient for practical calculations.

One option is to adopt the Curci-Ferrari (CF) model, a remarkably efficient framework to investigate the IR regime of YM theories~\cite{Pelaez21Rev}. Originally introduced in 1976~\cite{Curci76} as a renormalizable framework for massive vector bosons without the need of invoking a Higgs mechanism, the CF model underwent a revival in the last decade~\cite{Tissier10, Tissier11} emerging as a phenomenological proxy of a yet undiscovered IR completion of the FP action. 

The central idea is strikingly simple: the CF model consists in the simplest deformation of the FP action in the Landau gauge, including a phenomenologically motivated mass term for the gluon at tree-level, introduced after the gauge-fixing procedure. The result is a local and renormalizable extension of the FP action, with the same field content, which takes into account the dynamical mass generation observed in lattice simulations and recovers the standard FP results in the UV. The idea is to consider the CF action as a sensible starting point to study the IR regime of YM theories, understanding that the effective mass term already captures essential aspects of the IR dynamics such as the existence of Gribov copies. 

Despite its tremendous simplicity, the CF model shows a surprising ability to describe many important features of the IR regime, exhibiting good agreement with the results of lattice simulations. Indeed, it can be used, for instance, to compute correlations functions in YM theories with a very good agreement with lattice data already at one-loop order~\cite{Tissier10, Tissier11, Pelaez13, Barrios24a}. This success is not limited to the pure YM case, it can be extended to the inclusion of quark matter~\cite{Pelaez14, Pelaez15, Figueroa22}. Moreover, two-loop corrections improve the precision of the aforementioned results~\cite{Gracey19, Barrios20, Barrios21, Barrios22}. There is also a large number of promising results in the context of finite temperature, particularly regarding the deconfinement transition, as one can see, for instance, in Refs.~\cite{Reinosa14, Reinosa15a, Reinosa15b, Reinosa15c, Reinosa16, Reinosa17b}. Furthermore, it can be used to study spontaneous chiral symmetry breaking~\cite{Pelaez17, Pelaez21, Pelaez23}, to explore a new scenario for quark confinement~\cite{Pelaez25}, and to compute the non-Abelian Casimir energy~\cite{DeFabritiis26}.

The remarkable success of the CF model to describe the IR regime of YM theories using perturbation theory tools is deeply rooted in its renormalization group (RG) structure~\cite{Reinosa17, Oribe25}. A key feature is the existence of IR-safe trajectories, free from Landau poles, which ensure that the gauge coupling remains finite and moderate to all scales, allowing for a reliable perturbative expansion in the IR while keeping the well-known UV behavior. Moreover, this framework exhibits infrared fixed points which stabilize the RG flows. The RG structure of the CF model opens a perturbative window into the IR regime, a domain traditionally dominated by non‑perturbative effects, in sharp contrast with the standard FP results. 

Therefore, from everything presented so far, it is understood that the CF model has been successfully employed to describe the rich IR dynamics of pure YM theories and also of YM theories coupled with quark matter, which is the case of QCD. To this day, however, the CF model has never been used to investigate the IR dynamics of YM theories coupled with scalar matter. There are various reasons that make the study of such a coupled system interesting.

First of all, there are recent lattice studies dedicated to YM theories coupled with scalar matter~\cite{Maas19a, Maas19b, Maas16}, which allow us to access information from the IR regime through a genuinely non-perturbative method, comparing them with our perturbative analytical results. This offers us an excellent opportunity to test the CF framework, verifying its reliability beyond the already known cases. The coupling to scalar fields has another attractive feature: unlike fermions, scalars do not trigger spontaneous chiral symmetry breaking, thereby avoiding one of the most intricate non‑perturbative phenomena of QCD. This makes YM theories coupled with scalar matter a clean theoretical laboratory in which we can understand the influence of matter fields in the IR dynamics and investigate confinement in a simpler environment including matter.

We will pay special attention to the RG properties of this system. Given that a key feature for the success of the CF model is its RG structure, it is of utmost importance to investigate whether the most relevant properties of the RG flow are maintained in the scenario with scalar matter. In particular, it is important to ascertain whether infrared-safe trajectories exist in this expanded parameter space and whether novel fixed points could be found. Furthermore, the phase diagram of gauge theories coupled to scalar fields is known to be rich~\cite{Fradkin79}; therefore, taking the first steps in this topic from the CF perspective opens an interesting window for future investigation.

It is important to emphasize, however, that the results presented below must be taken {\it cum grano salis}. The issue is that, at the one-loop order of approximation considered in this work, and within the considered scheme, the two-point functions do not depend on the scalar self-coupling. As such, our computation is insensitive to this particular dimension of the parameter space. This could hide a potentially bad UV behavior coming from the scalar self-interactions, which could affect the present analysis and the RG properties presented here. Moreover, we will restrict ourselves to the symmetric potential scenario, not considering the possibility of a Higgs phase, which could bring interesting features. This work is a first foray in this system from the CF perspective; in the future, we plan to consider more general scenarios.  

The goal of this work is to investigate the CF model coupled to scalar matter and assess whether this framework can provide a successful description of this sector. In particular, we compute the scalar propagator at one-loop order and compare it with the lattice data reported in Ref.~\cite{Maas19a}. We also examine how the presence of scalar matter affects the gluon and ghost propagators. Finally, we analyze the RG flow structure at one-loop order for different scalar masses, identifying infrared-safe trajectories and novel fixed-points, while recovering the standard RG structure of the CF model in an appropriate limit.

This work is organized as follows. In Sec.~\ref{SecModel}, we present the CF model and its properties, introducing also the two-point functions that will be computed. Renormalization is discussed in Sec.~\ref{SecRenormalization}, while Sec.~\ref{SecRG} covers the renormalization group improvement, presenting the explicit one-loop expressions for the beta functions. The structure of the RG flow for different values of the scalar mass is discussed in Sec.~\ref{SecRGflow}. Our analytical results for the two-point functions are compared in Sec.~\ref{SecLattice} with available lattice data, and in Sec.~\ref{SecRGZ} with the RGZ results. Our concluding remarks are gathered in Sec.~\ref{SecConclusion}. In App.~\ref{AppFeynman}, we present the Feynman rules used in this work. A thorough discussion of the asymptotic behavior of the RG flow is provided in App.~\ref{app:asymp}.

\section{The model}\label{SecModel}
Let us briefly present the model considered in this work together with the main quantities to be computed.

\subsection{The Curci-Ferrari model}
In ${d=4-\epsilon}$  dimensions of Euclidean spacetime, the Landau gauge limit of the Curci-Ferrari (CF) model~\cite{Pelaez21Rev} is defined by the following Lagrangian:
	\begin{align}\label{CFlag}
		\mathcal{L}_{\rm CF} \!=\! \frac{1}{4} F_{\mu \nu}^a F_{\mu \nu}^a \!+\! i h^a \partial_\mu A_\mu^a + \partial_\mu \bar{c}^a (D_\mu c)^a \!+\! \frac{m^2}{2} A_\mu^a A_\mu^a\,,
	\end{align}
where $F_{\mu\nu}^a = \partial_\mu A_\nu^a - \partial_\nu A_\mu^a + g f^{abc} A_\mu^b A_\nu^c$ denotes the non-Abelian field strength tensor and $D_\mu^{ab} = \delta^{ab} \partial_\mu - g f^{abc} A_\mu^c$ is the covariant derivative in the adjoint representation, with $f^{abc}$ the SU(N) structure constants. The Faddeev-Popov (FP) ghosts are denoted by $\left(c^a, \bar{c}^a\right)$ and $h^a$ is the Nakanishi-Lautrup (NL) field used to enforce the Landau gauge condition ${\partial_\mu A_\mu^a=0}$. The model~\eqref{CFlag} has two parameters: the gauge coupling $g$ and the CF mass $m$.  

The tree-level gluon propagator associated to (\ref{CFlag}) is massive and transverse, that is
\begin{align}
    D_{\mu \nu}^{ab}(p) = \delta^{ab} \left(\frac{1}{p^2+m^2}\right) P^\perp_{\mu \nu}(p)\,,\label{eq:free}
\end{align}
with ${P^\perp_{\mu\nu}(p)=\delta_{\mu\nu}-p_\mu p_\nu/p^2}$. This, together with some symmetries to be discussed below, ensures that the model is renormalizable and thus predictive. Actually, the tree-level expression~\eqref{eq:free} has already the right qualitative behavior to match the lattice data for the Landau gauge gluon propagator in the IR regime, and, including one-loop corrections is already enough to obtain a very good agreement not only for the gluon, but also for the ghost propagator~\cite{Tissier10, Tissier11}, which improves even further upon including two-loop corrections~\cite{Gracey19}. The adequacy of the CF model to describe Landau gauge correlation functions is not limited to the quenched case; the results are also excellent in the presence of quark matter (see~\cite{Pelaez14, Barrios21}), although in this case, certain crucial quantities require one to go beyond perturbation theory (still within the CF model), see~\cite{Pelaez17, Pelaez21} and alike. It is not limited to correlation functions either, as it has been used to study various observables in the vacuum~\cite{Pelaez25, Pelaez23} or at finite temperature~\cite{Reinosa14, Reinosa15a, Reinosa15b, Reinosa15c, Reinosa16, Reinosa17}.

Our main goal in this work is to investigate whether such adequacy at describing YM theories in the IR extends in the presence of a coupling to (adjoint) scalar fields for which data are available in the literature~\cite{Maas19a}. We will remain at the simplest level of one-loop perturbation theory and leave refinements for future studies.

\subsection{Minimal coupling to scalar matter}
Let us then consider the CF model (\ref{CFlag}) minimally coupled to scalar matter in the adjoint representation, as described by the Lagrangian
\begin{align}
    \mathcal{L} = \mathcal{L}_{\rm CF} + \mathcal{L}_{\phi}\,,\label{eq:CFphi}
\end{align}
where the scalar sector is given by 
\begin{align}
    \mathcal{L}_{\phi} = \frac{1}{2} D_\mu \phi^a D_\mu \phi^a + \frac{m_\phi^2}{2} \phi^a \phi^a + \frac{\lambda}{4!}\left(\phi^a \phi^a\right)^2, 
\end{align}
with $\phi^a$ the adjoint scalar field. There are two extra parameters: the self-coupling $\lambda$ of the scalar field and the scalar mass $m_\phi$. In this work, we only consider positive values of $m^2_\phi$, that is, we do not consider the theory in the Higgs phase, while $\lambda$ is always considered positive for stability reasons.

The minimal coupling to scalar matter does not modify the pure gauge vertices, but introduces two new vertices representing the interaction of the scalar field with the gluon and a quartic self-interaction vertex for the scalar field. The Feynman rules can be found in App.~\ref{AppFeynman}.

\subsection{Symmetries}
The model (\ref{eq:CFphi}) enjoys a modified BRST symmetry which is the one already present in the CF model, appropriately extended to the scalar sector:
\begin{align}
s A_\mu^a &= D_\mu^{ab} c^b\,,\nonumber \\
s c^a &= -\frac{g}{2} f^{abc} c^b c^c\,,   \nonumber \\
s \bar{c}^a &= i h^a\,, \nonumber \\
s (i h^a) &= m^2 c^a\,,  \nonumber \\
s \phi^a &= - g f^{abc} c^b \phi^c.\label{eq:mBRST}
\end{align}
We will make use of some of the consequences of this symmetry below. 
\begin{figure}[t]
    \centering
    \includegraphics[width=0.5\textwidth]{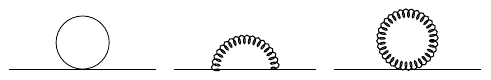}
    \caption{Feynman diagrams contributing to the scalar two-point vertex at one-loop order.}
    \label{FeynmanScalar}
\end{figure}

Alongside this symmetry, we stress that the transversality of the gluon propagator (\ref{eq:free}), together with the derivative nature of the ghost-antighost-gluon vertex (both consequences of the choice of Landau gauge), have also very useful consequences which still apply in the presence of scalar fields. Notably, beyond tree-level, each  contribution to a vertex function with external ghost and antighost legs comes naturally with one factor of external momentum per ghost or antighost leg. This implies that, as compared to the naive superficial degree of divergence, the actual degree of divergence of these vertex functions is decreased by as many units of mass dimension as there are external ghost and antighost legs. In particular, the ghost-antighost-gluon vertex function has no overall divergence, a fact which we will exploit below.

\begin{figure}[t]
\centering
\begin{minipage}[c]{0.3\linewidth}
\centering
\includegraphics[width=\linewidth,height=0.22\textheight,keepaspectratio]{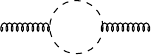}
\end{minipage}\hfill
\begin{minipage}[c]{0.3\linewidth}
\centering
\includegraphics[width=\linewidth,height=0.22\textheight,keepaspectratio]{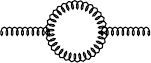}
\end{minipage}\hfill
\begin{minipage}[c]{0.3\linewidth}
\centering
\includegraphics[width=\linewidth,height=0.22\textheight,keepaspectratio]{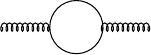}
\end{minipage}\hfill
\begin{minipage}[c]{0.3\linewidth}
\centering
\includegraphics[width=\linewidth,height=0.22\textheight,keepaspectratio]{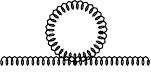}
\end{minipage}
\hspace{1 cm}
\begin{minipage}[c]{0.3\linewidth}
\centering
\includegraphics[width=\linewidth]{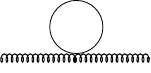}
\end{minipage}
\caption{Feynman diagrams contributing to the gluon two-point vertex at one-loop order.}
\label{FeynmanGluon}
\end{figure}

\begin{figure}[b]
    \centering
    \includegraphics[width=0.15\textwidth]{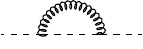}
    \caption{Feynman diagrams contributing to the ghost two-point vertex at one-loop order.}
     \label{FeynmanGhost}
\end{figure}

\subsection{Two-point correlation functions}

We aim to compute the two-point correlation functions for the scalar, gluon and ghost fields at one-loop order and compare our analytical results with available lattice data. More precisely, the scalar and gluon propagators and the ghost dressing function can be obtained as
\begin{align}
    G_\phi(p) = \frac{1}{\Gamma^{(2)}_\phi(p)}, \,\,     G_A(p) = \frac{1}{\Gamma^{(2)}_{A\perp}(p)}, \,\,   J_c(p) = \frac{p^2}{\Gamma^{(2)}_c(p)}\,,\label{eq:props}
\end{align}
that is by inverting the corresponding (1PI) two-point vertex functions which can be written symbolically as
\begin{align}
    \Gamma^{(2)}_{\phi}(p) &= p^2 + m^2_\phi - \Gamma_\phi^{\rm loop}(p)\,, \nonumber \\
    \Gamma^{(2)}_{A \perp}(p) &= p^2 + m^2 - \Gamma_{A \perp}^{\rm loop}(p)\,, \nonumber \\
    \Gamma^{(2)}_{c}(p) &= p^2 - \Gamma_c^{\rm loop}(p)\,,\label{eq:Gammas}
\end{align}
where we have left trivial color factors $\delta^{ab}$ implicit. 

With our sign convention, the loop contributions $\Gamma^{\rm loop}_X$  can be obtained by computing the one-loop Feynman diagrams depicted in Figs.~\ref{FeynmanScalar},~\ref{FeynmanGluon}, and~\ref{FeynmanGhost}. There are three diagrams contributing to the scalar two-point vertex; for the gluon, we have the three diagrams already present in the pure gauge case plus two new diagrams that arise due to the coupling of gluons to scalars; the ghost two-point vertex at one loop is computed through the same diagram as in the case without scalar matter since there is no new vertex appearing in this sector, but its final result will not be the same once renormalization group effects are incorporated since the scalar sector impacts the renormalization group flows of the model. Owing to the symmetry property referred to above,  $\Gamma_c^{\rm loop}(p)$ comes naturally with two factors of $p$, and owing to the IR regularization provided by the CF mass, the ghost dressing function $J_c(p)$ admits a finite limit as $\smash{p\to 0}$. $\Gamma_{A \perp}^{\rm loop}(p)$ refers to the transverse part, proportional to $P^\perp_{\mu \nu}$. 

The CF model also features a non-trivial longitudinal two-point gluon vertex $\Gamma^{(2)}_{A\parallel}(p)$. Although the latter does not enter the gluon propagator, it relates interestingly to the ghost dressing function. More precisely, because $\smash{\Gamma^{(2)}_{Ah}\neq 0}$, the inversion of the two-point vertex needs to be done in $(A,h)$-space. One finds the propagator
\begin{equation}
\delta^{ab}\left(
\begin{array}{cc}
\frac{P^\perp_{\mu\nu}(p)}{\Gamma^{(2)}_{A\perp}(p)}  & \frac{p_\mu}{p^2}\\\\
-\frac{p_\nu}{p^2} & \frac{\Gamma^{(2)}_{A\parallel}(p)}{p^2}
\end{array}
\right)\,,\label{eq:Ah_prop}
\end{equation}
where the first block in the diagonal is the gluon propagator while the second block is the propagator $\langle h^ah^b\rangle$ of the NL field. One can now exploit the modified BRST symmetry (\ref{eq:mBRST}) in the sector $(\bar c,ih)$ where it acts linearly, to relate the NL propagator to the ghost propagator. Indeed, expressing the invariance of the correlator $\langle\bar c(ih)\rangle$ under (\ref{eq:mBRST}) one finds
\begin{equation}
m^2\langle c \bar c\rangle(p)=\langle hh\rangle(p)\,,
\end{equation}
which, combined with (\ref{eq:props}) and (\ref{eq:Ah_prop}), leads to
\begin{equation}
\Gamma^{(2)}_{A\parallel}(p)=m^2J_c(p)\,.\label{eq:BRST}
\end{equation}
Note finally that $\Gamma^{(2)}_{A\parallel}(p)$ and $\Gamma^{(2)}_{A\perp}(p)$ should coincide in the $\smash{p\to 0}$ limit to ensure the regularity of the gluon two-point vertex, and thus
\begin{equation}
\Gamma^{(2)}_{A\perp}(p=0)=\Gamma^{(2)}_{A\parallel}(p=0)=m^2J_c(p=0)\,.\label{eq:BRST0}
\end{equation}

\section{Renormalization}\label{SecRenormalization}

\subsection{UV divergences}
The one-loop contributions to the two-point vertex functions diverge in four dimensions. Adopting dimensional regularization with ${d = 4- \epsilon}$, we find the divergences
\begin{align}
    \Gamma_\phi^{\rm loop}(p) \big\vert_{\rm div} &= \frac{18 g^2 N (p^2 + m^2) + \lambda m_\phi^2 (N^2+1)}{48 \pi^2}\frac{1}{\epsilon},\label{eq:div1}\\
    \Gamma^{\rm loop}_{A \perp}(p)\big\vert_{\rm div} &= \frac{g^2 N \left( 8p^2 - 3m^2\right)}{32 \pi^2 }\frac{1}{\epsilon},\label{eq:div2} \\
    \Gamma^{\rm loop}_c(p)\big\vert_{\rm div} &=  \frac{3 p^2 g^2 N }{32 \pi^2}\frac{1}{\epsilon}.\label{eq:div3}
\end{align}
A few remarks are in order at this stage, relating to the symmetries of the model discussed above. 

First of all, the divergence (\ref{eq:div3}) in the ghost two-point vertex is the same as in the absence of scalar field.  This is normal since the scalar field does not bring any contribution to the ghost self-energy at one-loop order. The divergence (\ref{eq:div2}) in the gluon two-point vertex is modified, but only in the term proportional to $p^2$. The term proportional to $m^2$ is not modified, which can be seen to be a consequence of (\ref{eq:BRST0}). As for the modification of the $p^2$-term, it has a direct impact on the coupling divergence of the model. Indeed, because the ghost-antighost-gluon vertex has no overall divergence (and thus no divergence at all at one-loop order), the coupling divergence can be read off from a proper combination of the gluon and ghost field divergences, obtained from the terms proportional to $p^2$ in (\ref{eq:div2})-(\ref{eq:div3}). More precisely, adding half the gluon field divergence to the ghost field divergence, one finds
\begin{equation}
\frac{g^2 N}{8 \pi^2 }\frac{1}{\epsilon}+\frac{3g^2 N }{32 \pi^2}\frac{1}{\epsilon}=\frac{7g^2 N }{32 \pi^2}\frac{1}{\epsilon}=x\frac{22}{3}\frac{g^2 N }{32 \pi^2}\frac{1}{\epsilon}\,,
\end{equation}
with $\smash{x=21/22}$. This is not anymore the well-known result in the absence of matter fields, which would correspond to $\smash{x=1}$, but we note that the model remains asymptotically free in the gauge sector.\footnote{As already emphasized before, to have a more complete picture one should take into account the running of $\lambda$, which could feature a UV Landau pole and change the results for some of the RG trajectories.}
Finally, the divergence (\ref{eq:div2}) does not involve any contribution proportional to $m_\phi^2$, despite the presence of two extra Feynman diagrams contributing to this vertex function as $\Delta\Gamma^{(2)}_{A\perp}$ and involving the scalar field. In fact, each of these diagrams has a divergence proportional to $m_\phi^2$ which cancel identically. The reason behind this cancellation is the standard BRST symmetry which is recovered in the $\smash{m\to 0}$ limit. Indeed, $\Delta\Gamma^{(2)}_{A\perp}$ is independent of $m$ and, then, without loss of generality, can be discussed in the limit $\smash{m=0}$. There, standard BRST symmetry implies that $\Gamma^{(2)}_A$ is transverse which means that it cannot have divergences proportional to $m_\phi^2$. This is true in the presence or absence of scalar fields, and because $\Delta\Gamma^{(2)}_{A\perp}$ appear as the difference of $\Gamma^{(2)}_A$ with or without scalar fields, the result follows. Actually, the very same argument allows one to show that $\smash{\Delta\Gamma^{(2)}_{A\perp}(p=0)=0}$, which we have verified using our one-loop results. 

Let us mention {\it en passant} that, as far as the IR structure is concerned, some of the most salient features of the model in the absence of scalar fields should be inherited by the model in the presence of the scalar field. This is because, at leading order, they rely exclusively on effective one-loop structures involving the massless ghost degrees of freedom. To cite just a few of these properties, the gluon propagator should still feature a non-monotonous behavior while the gluon three-point function should still display a zero-crossing at small momenta.

\subsection{Renormalization factors}

Everything we wrote down until here considered the bare Lagrangian, and thus must be rewritten in terms of renormalized objects to provide meaningful results, void of UV divergences. To this purpose, one introduces renormalization factors to express the bare fields and parameters in terms of renormalized ones:
\begin{align}
    A_B^{\mu, a} &= \sqrt{Z_A} A_R^{\mu, a}, \quad \,\, c_B^a = \sqrt{Z_c} c_R^{a}\,, \nonumber \\
    \phi_B^a &= \sqrt{Z_\phi} \phi_R^a\,, \quad \quad \bar{c}_B^a = \sqrt{Z_c} \bar{c}_R^{a}\,,
\end{align}
\begin{align}
    g_B &= Z_g g_R, \quad \, m_B^2 = Z_{m^2} m_R^2, \nonumber \\
    \lambda_B &= Z_\lambda \lambda_R, \quad m_{\phi B}^2 = Z_{m_\phi^2} m_{\phi R}^2,
\end{align}
Correspondingly, the bare two-point vertices are connected to their renormalized counterparts as
\begin{align}
    \Gamma^{(2)}_{R i} = Z_i \, \Gamma^{(2)}_{B i},\label{eq:GR}
\end{align}
where the index $i \in \{\phi, A, c\}$ changes depending on which two-point vertex is being considered. 

The renormalization factors are not all independent since the symmetries/structural properties of the model might constraint them. In the present case, there are two non-renormalization theorems stating that certain combinations of the renormalization factors are finite. First of all, the fact that the ghost-antighost-gluon vertex function is finite implies that the combination $Z_g Z_A^{1/2}Z_c$ is finite. Similarly, from Eq.~\eqref{eq:BRST}, which was a consequence of the modified BRST symmetry, one can also deduce that the combination $Z_{m^2}Z_AZ_c$ is finite.

At one-loop order, the renormalization factors can be written as $Z_i = 1 + \delta_i$, where the counterterms $\delta_i$ can come with a $g^2_R$ or a $\lambda_R$. Writing the renormalized two-point vertices in terms of the renormalized parameters, we can choose a suitable renormalization scheme to fix the counterterms and obtain a finite and meaningful result that could be compared to lattice data. From now on, it will be implicitly understood that all the quantities are renormalized; we will omit the subscripts to avoid making the notation cumbersome.

\subsection{Minimal subtraction}
The simplest renormalization scheme that can be considered is the minimal subtraction (MS) scheme, where the counterterms are chosen to just cancel the divergences, without bothering with the finite parts. This scheme allows us to easily find the renormalization factors needed to handle the divergences and give us relevant information about the UV behavior, which is universal at one-loop order. First, let us consider the scalar 1PI two-point function:
\begin{align}
    \Gamma_\phi^{(2)} = Z_\phi^{\rm MS} \left[p^2 + Z_{m_\phi^2}^{\rm MS} m_\phi^2 - \left(\frac{\Gamma_\phi^{\rm div}(p)}{\epsilon} + \Gamma_\phi^{\rm fin}(p)\right)\right], 
\end{align}
where we denoted $\Gamma_\phi^{\rm loop} = \frac{\Gamma_\phi^{\rm div}}{\epsilon} + \Gamma_\phi^{\rm fin}$ for simplicity. Writing the renormalization factors as  $Z_i^{\rm MS} = 1 + \delta_i^{\rm MS}$ and restricting to first order in the couplings, we get
\begin{align}
     \Gamma_\phi^{(2)}(p) &= p^2 + m_\phi^2 + \left[p^2 \delta_\phi^{\rm MS} + m_\phi^2\left(\delta_\phi^{\rm MS} + \delta_{m_\phi^2}^{\rm MS}\right)  -\Gamma_\phi^{\rm fin}(p)\right. \nonumber \\
     &\left. - \frac{1}{\epsilon}\left(\frac{18 g^2 N (p^2 + m^2) + \lambda m_\phi^2 (N^2+1)}{48 \pi^2} \right)\right].
\end{align}
In the above expression, since the loop correction $\Gamma^{\rm loop}_\phi$ is necessarily proportional to the couplings, at first order, we can simply substitute the bare parameters by the renormalized ones. Thus, we find for $Z_\phi^{\rm MS}$ and $Z_{m_\phi^2}^{\rm MS}$:
\begin{align}
    Z_\phi^{\rm MS} &= 1 +  \frac{3 g^2N}{8 \pi^2} \frac{1}{\epsilon}, \nonumber \\
    Z_{m_\phi^2}^{\rm MS} &= 1 + \left[\frac{3 g^2N}{8 \pi^2} \left(\frac{m^2}{m_{\phi}^2}-1\right) + \frac{\lambda \left(N^2+1\right)}{48 \pi^2}\right] \frac{1}{\epsilon}.
\end{align}

We can apply the same procedure for the two-point vertex functions in the gauge sector. Indeed, starting from the ghosts, we have at first order
\begin{align}
    \Gamma_c^{(2)}(p) = Z_c^{\rm MS} \left[p^2 - \frac{3 p^2 g^2 N }{32 \pi^2}\frac{1}{\epsilon} - \Gamma_c^{\rm fin}(p)\right].
\end{align}
Thus, in the MS scheme, we get immediately
\begin{align}
    Z_c^{\rm MS} = 1 + \frac{3 g^2 N}{32 \pi^2}\frac{1}{\epsilon}\,.
\end{align}
Now, for the gluon 1PI two-point function at first order,
\begin{align}
    \Gamma^{2}_{A \perp}(p) &= Z_A^{\rm MS} \left[ p^2 + Z_{m^2}^{\rm MS} m^2 \right. \nonumber \\
    &\left. - \frac{g^2 N \left( 8p^2 - 3m^2\right)}{32 \pi^2 }\frac{1}{\epsilon} - \Gamma^{\rm fin}_{A \perp}(p) \right], 
\end{align}
thus, in the MS scheme, one can readily find
\begin{align}
    Z_A^{\rm MS} &= 1 + \frac{g^2 N }{4 \pi^2 }\frac{1}{\epsilon}\,,  \nonumber \\
    Z_{m^2}^{\rm MS} &= 1 - \frac{ 11 g^2 N }{32 \pi^2 }\frac{1}{\epsilon}\,.
\end{align}

However, it is known that mass-independent schemes are generally not the most suitable for describing the IR regime in which we are mostly interested. Moreover, the MS scheme is known to be hampered by the presence of an IR Landau pole. Therefore, although very simple and useful for analytical manipulations, the MS scheme is not an appropriate scheme for our purposes; we need a suitable mass-dependent scheme, which we discuss now. 

\subsection{Infrared-safe scheme}
The pure gauge CF model has a distinguishing feature, it allows for a renormalization scheme in which perturbation theory is reliable at all scales, not presenting the Landau pole predicted by the FP Lagrangian, the so-called Infrared-safe (IRS) renormalization scheme. The IRS is defined by the following conditions:
\begin{align}\label{IRSscheme1}
    \Gamma^{(2)}_{A \perp}\left(p=\mu\right) = \mu^2 + m^2, \quad     \Gamma^{2}_c\left(p=\mu\right) = \mu^2,
\end{align}
supplemented by the two conditions
\begin{align}\label{IRSscheme2}
    Z_g \sqrt{Z_A} Z_c = 1\,,  \quad Z_{m^2} Z_A Z_c = 1\,,
\end{align}
which extends the two nonrenormalization theorems referred to above to the finite parts of the renormalization factors. The first condition means that the present scheme belongs to the family of Taylor schemes while the second condition can be seen, from Eq.~(\ref{eq:BRST}) and owing to the second condition in (\ref{IRSscheme1}), as a condition on the longitudinal part of the gluon two-point vertex, namely $\Gamma^{(2)}_{A\parallel}(p=\mu)=m^2$. 

We will adopt the IRS scheme in the gauge sector, Eqs.~\eqref{IRSscheme1} and~\eqref{IRSscheme2}, supplemented by a vanishing momentum (VM) scheme in the scalar sector, defined by the following renormalization conditions:
\begin{align} \label{ScalarRenormCond}
    \Gamma^{(2)}_\phi \left(p=\mu\right) &= \mu^2 + m_\phi^2\,, \nonumber \\
    \Gamma^{(2)}_\phi \left(p=0\right) &=  m_\phi^2\,.
\end{align}
Note that we are missing a renormalization condition to fix $Z_\lambda$. As we now explain, at our present level of approximation, this is because the renormalization of $\lambda$ is not needed.

Indeed, consider the renormalized scalar two-point vertex at one-loop order
\begin{align}
    \Gamma^{(2)}_\phi \! (p) \!=\! Z_\phi \! \left[p^2 \!+\! Z_{m_\phi^2} m_\phi^2 \!-\! \left(\lambda \Gamma_1 \!+\! g^2 \Gamma_2 (p) \!+\! g^2 \Gamma_3\right)\right],
\end{align}
where $\Gamma_i$ with $i=1,2,3$ refers to the Feynman diagrams in the order they appear in Fig.~\ref{FeynmanScalar}, with the corresponding coupling factorized. Due to their topology, the tadpole diagrams do not have a momentum dependence and are thus fully absorbed into the counterterms, owing to the above renormalization conditions. Since the $\lambda$-dependence occurs only via one tadpole, this means that the renormalized scalar two-point vertex will not depend on $\lambda$. More precisely, imposing the renormalization conditions~\eqref{ScalarRenormCond}, we get
\begin{align} \label{ScalarRenormCond2}
    Z_\phi \left[\mu^2 + Z_{m_\phi^2} m_\phi^2 - \left(\lambda \Gamma_1 + g^2 \Gamma_2(\mu) + g^2 \Gamma_3\right)\right] &= \mu^2 + m_\phi^2\,, \nonumber \\
    Z_\phi \left[ Z_{m_\phi^2} m_\phi^2 - \left(\lambda \Gamma_1 + g^2 \Gamma_2(0) + g^2 \Gamma_3\right)\right] &= m_\phi^2\,.
\end{align}
Writing $Z_i = 1 +\delta_i$, we find the following relation for the counterterms at first order in the couplings:
\begin{align}
    \delta_\phi &= \frac{g^2}{\mu^2}\left[\Gamma_2(\mu) - \Gamma_2(0)\right], \nonumber \\
    \left(\delta_\phi + \delta_{m_\phi^2}\right)m_\phi^2 &= \lambda \Gamma_1 + g^2 \Gamma_2(0) + g^2 \Gamma_3\,.
\end{align}
Thus,  the final result for the scalar 1PI two-point function at first order is
\begin{align}
    \Gamma^{(2)}_\phi(p) &= p^2 + m_{\phi}^2 \nonumber \\
    &- g^2 \left[\Gamma_2(p) - \frac{p^2}{\mu^2}\Gamma_2(\mu) - \left(1 - \frac{p^2}{\mu^2}\right) \Gamma_2(0)\right].
\end{align}
It is immediate to check that this expression indeed satisfies the renormalization conditions~\eqref{ScalarRenormCond} and that it is independent of $\lambda$, as announced. This feature will simplify the renormalization group analysis that will be done in the next section. However, one should always bear in mind that only considering the running of $\lambda$ one could tell whether perturbation theory is really well-grounded. We postpone this analysis to a future work.

\section{Renormalization Group Improvement} \label{SecRG}

The renormalization procedure introduces an arbitrary scale $\mu$, as one can see in the renormalization conditions in the previous section, for instance. This means that, even though our notation did not make it explicit, the renormalization factors and the renormalized quantities depend on this scale. In this section, we shall make this dependence explicit and exploit it. In particular, the renormalized two-point vertices $\Gamma^{(2)}_i(p,\mu)$ are both functions of the momentum and of the renormalization scale.

It is well known that the one-loop evaluation of the renormalized two-point vertex functions in the CF model is not accurate at large momenta $p$ if one considers a fixed scale $\mu$. This is because large logarithms appear in this limit that compete with the smallness of the gauge coupling. In the present model an additional complication occurs due to the fact that the scalar self-coupling is not necessarily small. To cope with the first problem, one can use one-loop perturbation theory to evaluate $\Gamma^{(2)}_i(p,\mu=p)$ which is void of large logarithms at large $p$ and use the renormalization group to reconstruct $\Gamma^{(2)}_i(p,\mu)$ from the knowledge of $\Gamma^{(2)}_i(p,\mu=p)$, so as to obtain RG-improved two-point vertices.\footnote{The last step is needed because the lattice does not evaluate $\Gamma^{(2)}_i(p,\mu=p)$, but rather $\Gamma^{(2)}_i(p,\mu)$ in some scheme at some fixed renormalization scale $\mu$. The connection between schemes/scales amounts to the multiplication by an appropriate constant.} In general this is done using the machinery of the Callan-Symanzik equation. In this case, however, one can shortcut the latter and proceed in a more direct way, as we now expose.

\subsection{RG-improved two-point vertices}
We first note that the two conditions (\ref{IRSscheme2}) can be used to express either the field renormalizations in terms of the parameter renormalizations, or the parameter renormalizations in terms of the field renormalizations. Let us here consider the first option which writes
\begin{equation}
Z_A=Z^2_g/Z^2_{m^2}\,, \quad Z_c=Z_{m^2}/Z_g^2\,.
\end{equation}
Then, taking the ratio of Eq.~\eqref{eq:GR} at two renormalization scales $\mu$ and $\mu=p$, and given that the bare quantities do not depend on the renormalization scale, one finds
\begin{eqnarray}
\frac{\Gamma^{(2)}_{A\perp}(p,\mu)}{\Gamma^{(2)}_{A\perp}(p,p)} & = & \frac{Z_A(\mu)}{Z_A(p)}=\frac{Z^2_g(\mu)}{Z^2_g(p)}\frac{Z^2_{m^2}(p)}{Z^2_{m^2}(\mu)}\,, \nonumber\\
\frac{\Gamma^{(2)}_{c}(p,\mu)}{\Gamma^{(2)}_{c}(p,p)} & = & \frac{Z_c(\mu)}{Z_c(p)}=\frac{Z_{m^2}(\mu)}{Z_{m^2}(p)}\frac{Z^2_{g}(p)}{Z^2_{g}(\mu)}\,.
\end{eqnarray}
Finally, upon exploiting that $Z_g \, g$ and $Z_{m^2}\, m^2$ do not depend on the renormalization scale, this rewrites
\begin{eqnarray}
\Gamma^{(2)}_{A\perp}(p,\mu) & = & \frac{m^4(\mu)}{m^4(p)}\frac{g^2(p)}{g^2(\mu)}\,\Gamma^{(2)}_{A\perp}(p,p)\,, \nonumber \\
\Gamma^{(2)}_{c}(p,\mu) & = & \frac{m^2(p)}{m^2(\mu)}\frac{g^2(\mu)}{g^2(p)}\,\Gamma^{(2)}_{c}(p,p)\,.
\end{eqnarray}

Since the renormalization conditions~\eqref{IRSscheme1} give $\Gamma^{(2)}_{A\perp}(p,p)=p^2+m^2(p)$ and $\Gamma^{(2)}_{c}(p,p)=p^2$, we can rewrite the renormalized vertex functions at a fixed scale $\mu$ solely in terms of the running parameters:
\begin{eqnarray}
\Gamma^{(2)}_{A\perp}(p,\mu) & = & \frac{m^4(\mu)}{g^2(\mu)}\frac{g^2(p)}{m^4(p)}\,(p^2+m^2(p))\,,\label{eq:gluon_imp} \nonumber \\
\Gamma^{(2)}_{c}(p,\mu) & = & \frac{g^2(\mu)}{m^2(\mu)}\frac{m^2(p)}{g^2(p)}\,p^2\,.\label{eq:ghost_imp}
\end{eqnarray}
We can proceed similarly for the scalar two-point vertex. Indeed, multiplying the second renormalization condition in Eq.~\eqref{ScalarRenormCond} by $Z_\phi^{-1}$ and re-expressing the renormalized scalar mass in terms of the bare one, we find
\begin{equation}
Z^{-1}_\phi(\mu)\Gamma^{(2)}_\phi(p=0,\mu)=\frac{Z_{m^2_\phi}(\mu)m^2_\phi(\mu)}{Z_{m^2_\phi}(\mu)Z_\phi(\mu)}\,,
\end{equation}
where the left-hand side is the bare two-point vertex at $p=0$ and should then not depend on $\mu$, just as the combination $Z_{m^2_\phi}(\mu)m^2_\phi(\mu)$ that appears in the numerator of the right-hand side. We deduce from this that the combination $Z_{m^2_\phi}(\mu)Z_\phi(\mu)$ does not depend on $\mu$, or, equivalently, that $Z_\phi(\mu)$ and $Z^{-1}_{m^2_\phi}(\mu)$ are the same up to a multiplicative constant.\footnote{We are not implying that $Z_{m^2_\phi}Z_\phi$ is finite, just that it does not depend on $\mu$. Thus, this is not a non-renormalization theorem in the sense discussed above.} Now, taking the ratio of Eq.~\eqref{eq:GR} at renormalization scales $\mu$ and $\mu=p$, we have
\begin{equation}
\frac{\Gamma^{(2)}_{\phi}(p,\mu)}{\Gamma^{(2)}_{\phi}(p,p)}=\frac{Z_\phi(\mu)}{Z_\phi(p)}=\frac{Z_{m^2_\phi}(p)}{Z_{m^2_\phi}(\mu)}\,.
\end{equation}
As before, exploiting that $Z_{m^2_\phi}(\mu)m^2_\phi(\mu)$ does not depend on the scale and that the first condition in Eq.~\eqref{ScalarRenormCond} rewrites $\Gamma^{(2)}_\phi(p,p)=p^2+m^2_\phi(p)$, we arrive at
\begin{equation}
\Gamma^{(2)}_{\phi}(p,\mu)=\frac{m^2_\phi(\mu)}{m^2_\phi(p)}(p^2+m^2_\phi(p))\,,\label{eq:scalar_imp}
\end{equation}
which expresses the renormalized scalar vertex function at a fixed scale in terms of the running scalar mass.

\subsection{Running parameters}
We are then left with the determination of the RG flow of the parameters $g(\mu)$, $m^2(\mu)$ and $m^2_\phi(\mu)$. In line with a remark made previously, the running of $\lambda$ does not enter the determination of the renormalized two-point vertices in the considered scheme.

The RG flow of the parameters is most easily obtained by considering small variations of the scale, defining the so-called beta functions
\begin{equation}
\beta_X\equiv \mu\frac{dX}{d\mu}\,,
\end{equation}
with $X \in \{g, m^2, m_\phi^2\}$. They are determined by noticing that the product $Z_XX$ of a renormalized quantity by the corresponding renormalization factor does not depend on the scale. Of course, this is also true for the logarithm $\ln (Z_XX)=\ln Z_X+\ln X$. Thus, one deduces that
\begin{equation}
0=\mu\frac{d\ln Z_X}{d\mu}+\mu\frac{d\ln X}{d\mu}\,,
\end{equation}
or, equivalently,
\begin{equation}
\beta_X=-X\gamma_X\,,
\end{equation}
with
\begin{equation}
\gamma_X\equiv\mu\frac{d\ln Z_X}{d\mu}\,,    
\end{equation}
the so-called anomalous dimensions. Thus, if the beta functions define the infinitesimal variation of the renormalized parameters with the renormalization scale, the anomalous dimensions provide the means for computing the beta functions from the renormalization factors.

From the field renormalization factors $Z_Y$, with $Y\in\{A,c,\phi\}$ one can also define corresponding anomalous dimensions $\gamma_Y$. A very convenient feature of the renormalization scheme that we are considering in this work is that the anomalous dimensions (and consequently the beta functions) associated to the parameters can be entirely expressed in terms of the anomalous dimensions (and consequently of the renormalized factors) associated to the fields. In particular, this enables one to access the running of $g$ without computing higher vertex functions.

Let us exploit once more the two renormalization conditions~\eqref{IRSscheme2} by expressing this time the renormalization factors associated to the parameters in terms of those associated to the fields:
\begin{equation}
Z_g=Z_A^{-1/2}Z_c^{-1}\,, \quad Z_{m^2}=Z_A^{-1}Z_c^{-1}\,.
\end{equation}
Taking the logarithms and differentiating with respect to the scale $\mu$, we arrive at
\begin{align}
\gamma_g &=-\left(\frac{\gamma_A}{2}+\gamma_c\right)\,, \nonumber \\
\gamma_{m^2} &=-(\gamma_A+\gamma_c)\,,
\end{align}
from which we can obtain
\begin{eqnarray}
    \beta_g & = & g \left(\frac{\gamma_A}{2} + \gamma_c\right), \label{eq:beta_g} \nonumber\\
    \beta_{m^2} & = & m^2 \left(\gamma_A + \gamma_c\right).\label{eq:beta_m2}
\end{eqnarray}
As for the running of $m^2_\phi$, recall that $Z^2_{m_\phi}Z_\phi$ does not depend on $\mu$ from which we deduce that
\begin{equation}
\gamma_{m^2_\phi}=-\gamma_\phi\,,
\end{equation}
and thus
\begin{equation}
\beta_{m^2_\phi}=m^2_\phi\gamma_\phi\,.\label{eq:beta_mphi2}
\end{equation}

To obtain the RG flows for the parameters $\{ g, m^2, m_\phi^2\}$, we need to solve the system of coupled differential equations  \eqref{eq:beta_m2} and \eqref{eq:beta_mphi2}, assuming some initial conditions $ g_0 \equiv g(\mu_0)$, $m^2_0 \equiv m^2(\mu_0)$, and $m^2_{\phi 0} \equiv m_\phi^2(\mu_0)$, defined at a renormalization scale $\mu_0$. 
Once the flows are obtained, the renormalized scalar and gluon propagators and the renormalized ghost dressing function at a fixed scale $\mu_0$ can be obtained, owing to Eqs.~\eqref{eq:props}, \eqref{eq:ghost_imp}, and \eqref{eq:scalar_imp} as
\begin{eqnarray} \label{FinalPropagRG}
G_\phi(p,\mu_0) & = & \frac{m^2_\phi(p)}{m^2_{\phi}(\mu_0)}\frac{1}{p^2+m^2_\phi(p)}\,, \nonumber \\
G_A(p,\mu_0) & = & \frac{m^4(p)}{m^4(\mu_0)}\frac{g^2(\mu_0)}{g^2(p)}\frac{1}{p^2+m^2(p)}\,, \nonumber\\
J_c(p,\mu_0) & = & \frac{m^2(\mu_0)}{m^2(p)}\frac{g^2(p)}{g^2(\mu_0)}\,.
\end{eqnarray}
The initial values will be chosen to minimize a suitable error function, achieving the best possible fit with the lattice data. This will be done in the next section.

\subsection{Explicit beta functions}
We here gather the explicit one-loop expressions for the beta functions and discuss interesting limits. We write the field anomalous dimensions as $\gamma_Y=\gamma^{\rm CF}_Y+\delta\gamma_Y$, where $\gamma^{\rm CF}_Y$ denotes the anomalous dimension in the CF model and $\delta\gamma_Y$ the correction coming from the scalar sector. 
The CF ones have already been computed in Ref.~\cite{Tissier11}:
\begin{eqnarray}
\gamma^{\rm CF}_A & \equiv & \zeta f_1(\tilde m^2)\nonumber\\
& = & \frac{\zeta}{6}\left[-17+\frac{74}{t}-\frac{12}{t^2}+t^2\ln t\right.\nonumber\\
& & \hspace{0.5cm}-\,\frac{(t-2)^2(2t-3)(t+1)^2}{t^3}\ln(t+1)\nonumber\\
& & \hspace{0.5cm}\left.-\frac{\sqrt{t+4}}{t^{3/2}}(t^3-9t^2+20t-36)\ln\frac{\sqrt{t+4}-\sqrt{t}}{\sqrt{t+4}+\sqrt{t}}\right],\nonumber\\
\gamma^{\rm CF}_c & \equiv & \zeta f_2(\tilde m^2)\nonumber\\
& = & -\frac{\zeta}{2}\left[2+\frac{2}{t}-t\ln t+\frac{(t+1)^2(t-2)}{t^2}\ln(t+1)\right],\nonumber\\
\gamma^{\rm CF}_\phi & = & 0\,,
\end{eqnarray} 
where we introduced $\zeta\equiv g^2 N/(16\pi^2)$ and $t\equiv\mu^2/m^2\equiv 1/\tilde m^2$. The corrections from the scalar sector are:
\begin{align}
\delta\gamma_A &\equiv  \zeta f_3(\tilde m^2_\phi)\nonumber\\
&=  \frac{\zeta}{3}\left[1+\frac{12}{s}+6\frac{\sqrt{4+s}}{s^{3/2}}\ln\frac{\sqrt{4+s}-\sqrt{s}}{\sqrt{4+s}+\sqrt{s}}\right],\nonumber\\
\delta\gamma_c &=  0\,,\nonumber\\
\delta\gamma_\phi &\equiv  \zeta f_4(\tilde m^2,\tilde m^2_\phi)\nonumber\\
&=  \frac{\zeta}{\tilde m^2}\bigg\lbrace -4\tilde m^2 \left(1+\tilde m^2 - 2 \tilde m_\phi^2\right) -  \left[1 + 2 \left(\tilde m^2 - \tilde m_\phi^2\right)^3 \right.\nonumber \\
&\left. + 3 \left(\tilde m^2 - \tilde m_\phi^2\right)\left(\tilde m^2 + \tilde m_\phi^2\right)\right] \log\left(\frac{\tilde m^2}{\tilde m_\phi^2}\right) \nonumber \\
&+2 A_+ A_- \left[ \tilde m^2 +\tilde m_\phi^2 - 1 +2 \left(\tilde m^2 - \tilde m_\phi^2\right)^2 \right] \log\left(\frac{A_+ + A_-}{A_+ - A_-}\right) \nonumber \\ 
&-2 \left(1 + \tilde m_\phi^2\right)^2\left(2 \tilde m_\phi^2-1\right) \log\left(\frac{1+\tilde m_\phi^2}{\tilde m_\phi^2}\right)  \bigg\rbrace\,,
\end{align} 
where we have introduced $s\equiv\mu^2/m^2_\phi\equiv 1/\tilde m^2_\phi$ as well as
\begin{align}
A_\pm \equiv \left[1 + \left(\tilde m \pm \tilde m_\phi \right)^2\right]^{1/2}\,.
\end{align}
The functions $f_1(\tilde m^2)$, $f_2(\tilde m^2)$, $f_3(\tilde m^2_\phi)$, $f_4(\tilde m^2,\tilde m^2_\phi)$ have been introduced for later purpose. The first two are monotonically increasing functions of $\tilde m^2$, with boundary values
\begin{eqnarray}
f_1(0) & = & -\frac{13}{3}\,, \quad f_1(\infty)=\frac{1}{3}\,,\nonumber\\
f_2(0) & = & -\frac{3}{2}\,, \quad f_2(\infty)=0\,.\label{eq:limits1}
\end{eqnarray}
Similarly, $f_3(\tilde m^2_\phi)$ is a monotonically decreasing function of $\tilde m^2_\phi$, with boundary values
\begin{eqnarray}
f_3(0) & = & \frac{1}{3}\,, \quad f_3(\infty)=0\,.\label{eq:limits2}
\end{eqnarray}
Finally, $f_4(\tilde m^2,\tilde m^2_\phi)$ is a monotonically increasing function of both $\tilde m^2$ and $\tilde m^2_\phi$, with boundary values
\begin{eqnarray}
f_4(0^+,\tilde m^2_\phi) & < & \infty\,, \quad f_4(\tilde m^2\gg\tilde m^2_\phi,\tilde m^2_\phi)\sim -2\frac{\ln\tilde m^2}{\tilde m^2}\,,\nonumber\\
f_4(\tilde m^2,0^+) & < & \infty\,, \quad f_4(\tilde m^2,\tilde m^2_\phi\gg\tilde m^2)\sim -\frac{4}{\tilde m^2_\phi}\,,\label{eq:limits3}
\end{eqnarray}
and
\begin{equation}
f_4(0^+,0^+)=-6\,.
\end{equation}
In particular, for strictly finite values of $\tilde m^2$ and $\tilde m^2_\phi$, ${f_4(\tilde m^2,\tilde m^2_\phi)<0}$. Although we are not able to present a formal proof of these features of $f_4(\tilde m^2,\tilde m^2_\phi)$, this was the conclusion reached after making  extensive numerical tests using {\it Mathematica}.

\section{structure of the RG flow} \label{SecRGflow}

To study the structure of the flow, it is convenient to consider the flow equations for the dimensionless quantities $\zeta$, $\tilde m^2$ and $\tilde m^2_\phi$. Using that
\begin{equation}
\frac{\beta_\zeta}{\zeta}=2\frac{\beta_g}{g}\,, \quad \frac{\beta_{\tilde M^2}}{\tilde M^2}=\frac{\beta_{M^2}}{M^2}-2\,,
\end{equation}
where $M = \{m , m_\phi\}$, we find
\begin{eqnarray}
\beta_\zeta & = & \zeta(\gamma_A+2\gamma_c)\,,\nonumber\\
\beta_{\tilde m^2} & = & \tilde m^2(\gamma_A+\gamma_c-2)\,,\nonumber\\
\beta_{\tilde m_\phi^2} & = & \tilde m^2_\phi(\gamma_\phi-2)\,.\label{eq:flows}
\end{eqnarray}
It follows that the flow equations have the structure
\begin{eqnarray}
\beta_\zeta & = & \zeta^2[f_1(\tilde m^2)+2f_2(\tilde m^2)+f_3(\tilde m^2_\phi)]\,,\nonumber\\
\beta_{\tilde m^2} & = & \tilde m^2[\zeta(f_1(\tilde m^2)+f_2(\tilde m^2)+f_3(\tilde m^2_\phi))-2]\,,\nonumber\\
\beta_{\tilde m_\phi^2} & = & \tilde m^2_\phi[\zeta f_4(\tilde m^2,\tilde m^2_\phi)-2]\,.\label{eq:flow_struc}
\end{eqnarray}
Furthermore, the flow equation for $\tilde m^{-2}_\phi \equiv s$ writes
\begin{equation}
\beta_{\tilde m^{-2}_\phi}=\tilde m^{-2}_\phi[2-\zeta f_4(\tilde m^2,\tilde m^2_\phi)]\,.
\end{equation}
Since $f_4(\tilde m^2,\tilde m^2_\phi)$ is strictly negative for finite values of $\tilde m^2$ and $\tilde m^2_\phi$, we deduce that the only possible fixed-points must have $\tilde m^2_\phi=\infty$ or ${\tilde m^2_\phi=0}$. In fact, the planes ${\tilde m^2_\phi=\infty}$ and ${\tilde m^2_\phi=0}$ in the parameter space are stable in the sense that any RG trajectory starting on them will remain there because ${\beta_{\tilde{m}_\phi^2}=0}$. Before studying the RG flow structure in the bulk of the three-dimensional parameter space $(\tilde m^2,\tilde m^2_\phi,\zeta)$, we can consider what happens in the RG-invariant planes ${\tilde m^2_\phi=\infty}$ and ${\tilde m^2_\phi=0}$ (while restricting to positive values of $\tilde m^2$ and $\zeta$).

\begin{figure}[t!]
		\begin{minipage}[b]{0.8\linewidth}
			\includegraphics[width=\textwidth]{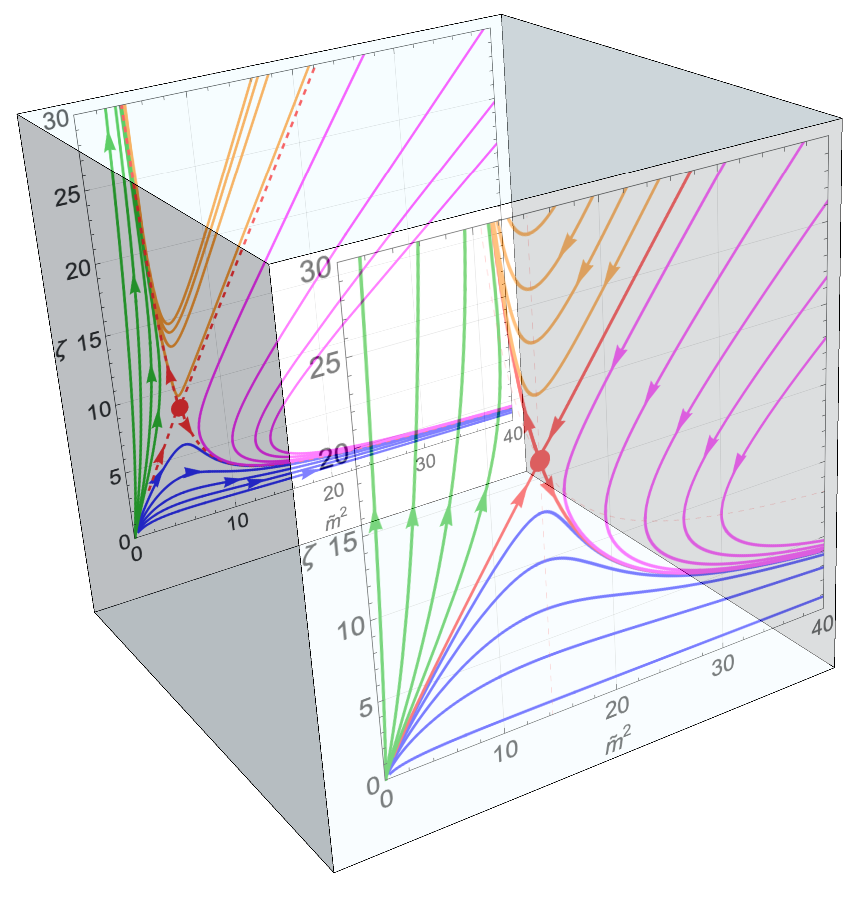}
		\end{minipage}  \hfill
		\caption{Renormalization group flow in the plane ${\tilde m^2_\phi=\infty}$ (front), and in the plane ${\tilde m^2_\phi=0}$ (back). The structure of the flow in the bulk ${0<\tilde m^2_\phi<\infty}$ is discussed in the main text.}\label{fig:flowA} 
		\label{fig:Cylinder}
\end{figure}

\subsection{RG flow analysis on the boundary ${\tilde m^2_\phi=\infty}$}

In the plane ${\tilde m^2_\phi=\infty}$ and owing to the limits (\ref{eq:limits1})-(\ref{eq:limits2}), the flow equations boil down to the ones in the pure CF model:
\begin{eqnarray}
\beta_\zeta & = & \zeta^2[f_1(\tilde m^2)+2f_2(\tilde m^2)]\,,\nonumber\\
\beta_{\tilde m^2} & = & \tilde m^2[\zeta(f_1(\tilde m^2)+f_2(\tilde m^2))-2]\,.
\end{eqnarray}
This is expected since, for any finite $\mu$, ${\tilde m^2_\phi=\infty}$ corresponds to ${m^2_\phi=\infty}$ in which case we expect the scalar field to decouple. The corresponding flow has been studied in Refs.~\cite{Reinosa17, Gracey19, Oribe25}: the plane $(\tilde m^2,\zeta)$ is subdivided in four different regions, corresponding to four different types of theories, see Fig.~\ref{fig:flowA}. Three of them contain an UV or an IR Landau pole (or both) and the fourth is asymptotically free both in the UV and in the IR. For this type of theories, the UV corresponds to ${\tilde m^2\to 0}$ and the IR to ${\tilde m^2\to\infty}$. Owing to (\ref{eq:limits1})-(\ref{eq:limits2}), the beta functions become
\begin{equation}
\beta_\zeta^{\rm UV}=-\frac{22}{3}\zeta^2\,, \quad \beta^{\rm UV}_{m^2}= -\frac{35}{6}\zeta m^2\,,\label{eq:betauvinf}\\
\end{equation}
in the UV. The first equation rewrites $d(1/\zeta)/d\ln\mu=22/3$ which integrates to
\begin{equation}
\frac{1}{\zeta(\mu)}=\frac{1}{\zeta(\mu_0)}+\frac{22}{3}\ln\left(\frac{\mu}{\mu_0}\right)\,,
\end{equation}
while the ratio of the two beta functions rewrites $d\ln m^2/d\ln \zeta=35/44$ which integrates to
\begin{equation}
m^2(\mu)=m^2(\mu_0)\left(\frac{\zeta(\mu)}{\zeta(\mu_0)}\right)^{35/44}\,.\label{eq:62}
\end{equation}
Although correct, this identity hides a subtlety related to the fact that we are actually integrating asymptotic relations. In particular, exponentiation of such relations, which we have used here, usually requires a proper check which we discuss in App.~\ref{app:asymp}. Similarly, the beta functions in the IR become
\begin{equation}
\beta_\zeta^{\rm IR}=\frac{1}{3}\zeta^2\,, \quad \quad \,\, \beta^{\rm IR}_{m^2}= \frac{1}{3}\zeta m^2\,,\label{eq:betairinf}
\end{equation}
from which it follows that
\begin{equation}
\frac{1}{\zeta(\mu)}=\frac{1}{\zeta(\mu_0)}-\frac{1}{3}\ln\left(\frac{\mu}{\mu_0}\right)\,,\label{eq:71}
\end{equation}
and
\begin{equation}
m^2(\mu)=m^2(\mu_0)\frac{\zeta(\mu)}{\zeta(\mu_0)}\,,\label{eq:72}
\end{equation}
see App.~\ref{app:asymp} for more details.

The coupling $g$ and the dimensionful mass $m$ go to zero in the UV and in the IR. Note finally that the four families of RG trajectories are separated by two separatrix curves that meet at a fixed point $(\tilde m^2_*,\zeta_*)$, such that 
\begin{eqnarray}
0 & = & f_1(\tilde m^2_*)+2f_2(\tilde m^2_\star)\,,\nonumber\\
\zeta_* & = & \frac{2}{f_1(\tilde m^2_*)+f_2(\tilde m^2_\star)}\,.
\end{eqnarray}
From the above equations, one finds
\begin{equation}
\tilde m^2_*\simeq 14.17\,, \quad \zeta_*\simeq 16.11\,,
\end{equation}
as already reported in Ref.~\cite{Oribe25}.

\subsection{RG flow analysis on the boundary ${\tilde m^2_\phi=0}$}

In the plane ${\tilde m^2_\phi=0}$, instead, and owing to the limits~\eqref{eq:limits1}-\eqref{eq:limits2}, the flow equations become:
\begin{eqnarray}
\beta_\zeta & = & \zeta^2\left[f_1(\tilde m^2)+2f_2(\tilde m^2)+1/3\right]\,,\\
\beta_{\tilde m^2} & = & \tilde m^2\left[\zeta(f_1(\tilde m^2)+f_2(\tilde m^2)+1/3)-2\right]\,.
\end{eqnarray}
The corresponding flow can be studied following Ref.~\cite{Oribe25}: the plane $(\tilde m^2,\zeta)$ is again subdivided in four different regions, corresponding to four different types of theories, see Fig.~\ref{fig:flowA}. Three of them contain an UV or an IR Landau pole (or both) and the fourth is  asymptotically free both in the UV and in the IR. For this type of theories, the UV corresponds to $\tilde m^2\to 0$ and the IR to $\tilde m^2\to\infty$. Owing to Eq.~\eqref{eq:limits1}-\eqref{eq:limits2}, the beta functions become
\begin{equation}
\beta_\zeta^{\rm UV}=-7\zeta^2\,, \quad \beta^{\rm UV}_{m^2}= -\frac{11}{2}\zeta m^2\,,\label{eq:betauv0}\\
\end{equation}
in the UV, from which it follows that
\begin{equation}
\frac{1}{\zeta(\mu)}=\frac{1}{\zeta(\mu_0)}+7\ln\left(\frac{\mu}{\mu_0}\right)\,,\label{eq:78}\\
\end{equation}
and
\begin{equation}
m^2(\mu)=m^2(\mu_0)\left(\frac{\zeta(\mu)}{\zeta(\mu_0)}\right)^{11/14},\label{eq:79}
\end{equation}
while
\begin{equation}
\beta_\zeta^{\rm IR}=\frac{2}{3}\zeta^2\,, \quad \,\,\,\, \beta^{\rm IR}_{m^2}= \frac{2}{3}\zeta m^2\,,\label{eq:betair0} 
\end{equation}
in the IR, from which it follows that 
\begin{equation}
\frac{1}{\zeta(\mu)}=\frac{1}{\zeta(\mu_0)}-\frac{2}{3}\ln\left(\frac{\mu}{\mu_0}\right)\,,\\
\end{equation}
and
\begin{equation}
m^2(\mu)=m^2(\mu_0)\frac{\zeta(\mu)}{\zeta(\mu_0)}\,,
\end{equation}
where, as before, we refer to App.~\ref{app:asymp} for more details.

Once again, the coupling $g$ and the dimensionful mass $m$ go to zero in the UV and in the IR. Moreover, the four families of RG trajectories are separated by two separatrix curves that meet at a fixed point $(\tilde m^2_*,\zeta_*)$ such that
\begin{eqnarray}
0 & = & f_1(\tilde m^2_*)+2f_2(\tilde m^2_\star)+1/3\,,\nonumber\\
\zeta_* & = & \frac{2}{f_1(\tilde m^2_*)+f_2(\tilde m^2_\star)+1/3}\,.
\end{eqnarray}
From the above equations, one can find
\begin{equation}
\tilde m^2_*\simeq 5.78\,, \quad \zeta_*\simeq 8.66\,,
\end{equation}
which is a novel fixed point. However, we should pay attention that this fixed point is found far beyond the range of applicability of the perturbative expansion used here, corresponding to ${\zeta\ll 1}$, just as in the pure CF case.\footnote{In the latter case, two-loop corrections move the fixed point to values $\zeta$ closer to $1$, although still above, see Ref.~\cite{Gracey19}.}

\begin{figure}[t!]
		\begin{minipage}[b]{1.1\linewidth}
			\includegraphics[width=\textwidth]{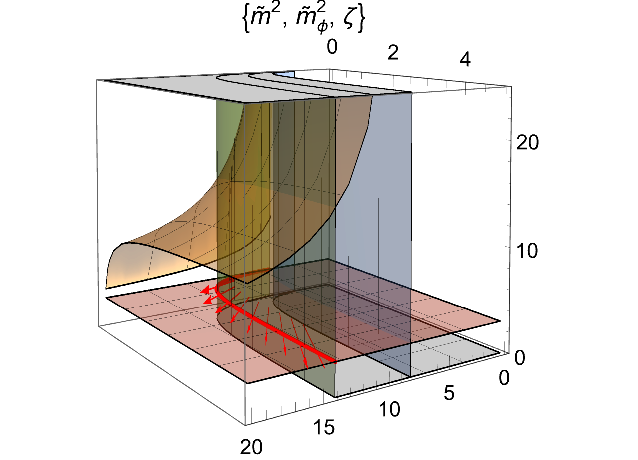}
		\end{minipage}
		\caption{Hypersurfaces H1 (green) and H2 (orange) together with the plane ${\zeta=3}$ (red) and the vertical asymptote surface to H2 (blue) in parameter space with ${\tilde m^2\in[0,20]\,}$, ${\tilde m^2_\phi\in[0,5]}$ and ${\zeta\in [0,25]}$. The red curve represents a slice of hypersurface H1 at ${\zeta=3}$ and the arrows indicate the direction of the flow on this curve, from the UV to the IR.}
		\label{HypersBulkScalarCF}
\end{figure}

\subsection{RG flow analysis in the bulk $\left(0 < \tilde{m}^2_\phi < \infty\right)$}

So far, we considered the flow structure within the boundary planes ${\tilde m^2_\phi=\infty}$ and ${\tilde m^2_\phi=0}$ that are stable under the renormalization group. To study the structure of the flow inside the bulk ${0<\tilde m^2_\phi<\infty}$, we go back to Eq.~\eqref{eq:flow_struc} and divide the parameter space $(\tilde m^2,\tilde m^2_\phi,\zeta)$ according to the hypersurfaces H1 and H2, defined by:
\begin{align}
{\rm H1: \quad} 0  =  f_1(\tilde m^2)+2f_2(\tilde m^2)+f_3(\tilde m^2_\phi) \label{eq:hyp1}
\end{align}
and
\begin{align}
{{\rm H2}: \quad} \zeta  =  \frac{2}{f_1(\tilde m^2)+f_2(\tilde m^2)+f_3(\tilde m^2_\phi)} \; \wedge \; \zeta > 0\,.\label{eq:hyp2}
\end{align}
On these hypersurfaces, we have ${\beta_\zeta=0}$ and ${\beta_{\tilde m^2}=0}$. They are useful because they define regions in the parameter space where the signs of $\beta_\zeta$ and $\beta_{\tilde m^2}$ are fixed, thus determining the directions of the flow along the $\zeta$- and $\tilde m^2$-axes. We recall that $\beta_{\tilde m^2_\phi}$ is always strictly negative in the bulk, meaning that the flow along the $\tilde m^2_\phi$-axis is always towards increasing values of $\tilde m^2_\phi$ as ones decreases the renormalization scale down to the IR.

It will be convenient to label the different regions. To do so, we first note that the hypersurface H1 is vertical along the $\zeta$ direction (indeed, its defining equation is $\zeta$-independent). Then, among the two regions in which this hypersurface splits the parameter space, one contains the whole $\zeta$-axis ${(\tilde m^2,\tilde m^2_\phi)=(0,0)}$. We refer to this region as the lower region associated the hypersurface H1, while the other region is the upper one. Similarly, the hypersurface H2 can be seen to subdivide the parameter space into a region that contains the whole plane ${\zeta=0}$, which we refer to as the lower region associated to the hypersurface H2, while the other region is referred to as the upper one.  For latter purpose, we can be slightly more precise about the hypersurface H2. First of all, because $f_1(\tilde m^2)$ and $f_2(\tilde m^2)$ are monotonically increasing while $f_3(\tilde m^2_\phi
)$ is monotonically decreasing, we have $f_1(\tilde m^2)+f_2(\tilde m^2)+f_3(\tilde m^2_\phi)\leq f_1(\infty)+f_2(\infty)+f_3(0)=1/3+0+1/3=2/3$. It follows that the hypersurface H2 is bounded from below by the plane ${\zeta=3}$. Second, H2 presents an asymptotic hypersurface $0=f_1(\tilde m^2)+f_2(\tilde m^2)+f_3(\tilde m^2_\phi)$, the approach of which makes $
\zeta$ diverges positively. This hypersurface is also vertical, as H1, and belongs to the lower region associated to H1, see Fig.~\ref{HypersBulkScalarCF}, since then $f_1(\tilde m^2)+2f_2(\tilde m^2)+f_3(\tilde m^2_\phi)=f_2(\tilde m^2)<0$, just as $f_1(0)+2f_2(0)+f_3(0)= -7 <0$. 

We have thus four regions ${\cal R}_{\uparrow\uparrow}$, ${\cal R}_{\uparrow\downarrow}$, ${\cal R}_{\downarrow\uparrow}$, ${\cal R}_{\downarrow\downarrow}$, where, for instance, ${\cal R}_{\uparrow\downarrow}$ means the intersection between the upper region associated to the hypersurface H1 and the lower region associated to the hypersurface H2. As already mentioned, each of these regions is characterized by fixed signs for $\beta_\zeta$ and $\beta_{\tilde m^2}$: the first arrow pointing up (resp. down) corresponds to a positive (resp. negative) $\beta_\zeta$, while the second arrow pointing up (resp. down) corresponds to a positive (resp. negative) $\beta_{\tilde m^2}$. To conclude this, one can for instance use Eq.~\eqref{eq:flow_struc} together with Eqs.~\eqref{eq:limits1}-\eqref{eq:limits3} to see that, along the $\zeta$-axis ${(\tilde m^2,\tilde m^2_\phi)=(0,0)}$, one has ${\beta_\zeta = \zeta^2 [-13/3 + 2(-3/2) + 1/3] < 0}$, while in the plane ${\zeta=0}$, one has $\beta_{\tilde{m}^2} = \tilde{m}^2 \left[0(f_1+f_2+f_3) -2\right] < 0$.

We stress that the  regions thus defined are not regions to which the flow is restricted, since the RG trajectories can cross the hypersurfaces. However, when they do so, they do it in a very specific way that strongly constrains the RG trajectories. Indeed, at any crossing of H1, the flow is orthogonal to the $\zeta$ direction (since ${\beta_\zeta=0}$), while at any crossing of H2, the flow is orthogonal to the $\tilde m^2$ direction (since ${\beta_{\tilde m^2}=0}$). Using these remarks, one can try to analyze the different families of RG trajectories that make the flow diagram, along the same lines as was done in Ref.~\cite{Oribe25}. The general analysis is more difficult here, however, due to the additional dimension $\tilde m^2_\phi$. So, in the following, we restrict the analysis to a particular class of trajectories that fit our needs.

\begin{figure}[t!]
		\begin{minipage}[b]{0.9\linewidth}
			\includegraphics[width=\textwidth]{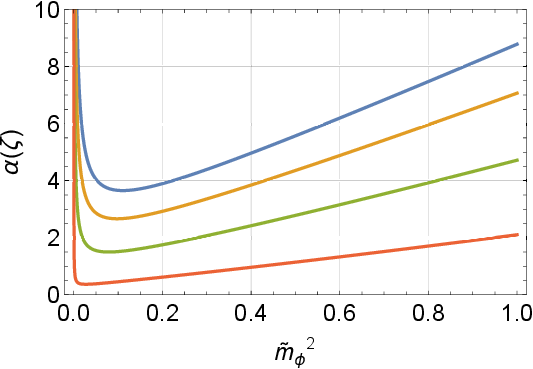}
		\end{minipage}  \hfill
		\caption{The ratio $\alpha(\zeta)$ along the hypersurface H1~\eqref{eq:hyp1} for various values of $\zeta$. From top to bottom: $\zeta=1$, $2$, $4$ and $8$.}
		\label{fig:alpha}
\end{figure}

Suppose now that we initialize the flow at a point $(\tilde m^2_0,\tilde m^2_{\phi0},\zeta_0)$ that lies on the hypersurface H1, with ${\zeta\leq 3}$, and so, at the boundary between regions ${\cal R}_{\downarrow\downarrow}$ and ${\cal R}_{\uparrow\downarrow}$.  At least for ${\zeta\leq 3}$, the flow trajectories that pierce the hypersurface H1, do it from region ${\cal R}_{\downarrow\downarrow}$ to region ${\cal R}_{\uparrow\downarrow}$ as $\mu$ is decreased. To check this, we evaluate the ratio $\alpha(\zeta)$ between the direction of the flow $\beta_{\tilde m^2}/\beta_{\tilde m^2_\phi}$ and that, $-f'_3(\tilde m^2_\phi)/(f'_1(\tilde m^2)+2f'_2(\tilde m^2))$, of the tangent along the hypersurface H1 at a given value of $\zeta$. In Fig.~\ref{fig:alpha}, we illustrate that this ratio is larger than $1$ when ${\zeta\leq 3}$ and thus that the piercing of the hypersurface H1 is done as mentioned, see Fig.~\ref{HypersBulkScalarCF} for an illustration at height ${\zeta=3}$. 

As we decrease $\mu$, the flow trajectory enters region ${\cal R}_{\uparrow\downarrow}$ where $\zeta$ decreases while $\tilde m^2$ increases. But then the flow cannot cross back to ${\cal R}_{\downarrow\downarrow}$ since, if it did, it would do it with a smaller value of $\zeta$ for which it is even more true that the piercing of the hypersurface H1 should be as in Fig.~\ref{HypersBulkScalarCF} and not in the opposite direction. The flow cannot cross to region ${\cal R}_{\uparrow\uparrow}$ since the hypersurface H2 is bounded from below by ${\zeta=3}$. In summary, the flow remains in region $\cal{R}_{\uparrow\downarrow}$ as $\mu$ is decreased further down to ${\mu\to 0}$. The same conclusion applies to the flow as $\mu$ is increased from the considered point in the hypersurface H1: it enters region ${\cal R}_{\downarrow\downarrow}$ and remains there up to ${\mu\to\infty}$. These trajectories are characterized by a bounded $\zeta$, thus corresponding to infrared-safe trajectories. Let us now analyze more precisely the behavior of these infrared-safe trajectories in the deep IR (${\mu\to 0}$) and the deep UV (${\mu\to\infty}$).

As one increases $\mu$ from the chosen initial point on H1, $\zeta$, $\tilde m^2$ and $\tilde m^2_\phi$ decrease and, thus, admit a limit as ${\mu\to\infty}$. This implies that, in this limit, $\beta^{\rm UV}_\zeta\simeq C_{\rm UV} \zeta^2$ with $C_{\rm UV}<0$ (since the flow occurs in region ${\cal R}_{\downarrow\downarrow}$). Then, 
\begin{equation}
\zeta(\mu)=\frac{\zeta_0}{1-C_{\rm UV}\zeta_0 \ln\left(\frac{\mu}{\mu_0}\right)}\,,\label{eq:zeta_uv}
\end{equation}
which approaches $0^+$ as ${\mu\to\infty}$. It then follows that the beta function of $\tilde m^2$ is dominated by $-2\tilde m^2$, and then 
\begin{equation}
\ln \tilde m^2/\tilde m_0^2= \ln \mu_0^2/\mu^2\,,\label{eq:log}
\end{equation}
which is only possible if ${\tilde m^2\to 0}$. Following a similar reasoning, we find that ${\tilde m^2_\phi\to 0}$ and, thus, the beta functions for $\zeta$, $\tilde m^2$ and $\tilde m^2_\phi$ then become those in Eq.~\eqref{eq:betauv0}, fixing ${C_{\rm UV}=-7}$, and
\begin{equation}
\beta^{\rm UV}_{m_\phi^2}=-6\zeta m^2_\phi\,.
\end{equation}
This implies (\ref{eq:78})-(\ref{eq:79}) and\footnote{Once again, deriving this formula properly requires some work, along the lines of App.~\ref{app:asymp}.}
\begin{equation}
m^2_\phi(\mu)=m^2_\phi(\mu_0)\left(\frac{\zeta(\mu)}{\zeta(\mu_0)}\right)^{6/7}\,,\label{eq:911}
\end{equation}
which, in turn, shows that the dimensionful scalar mass goes to $0$ in the UV. So, when flowing to the UV from the considered initial condition, the flow is attracted towards the asymptotically free UV fixed point that lies in the ${\tilde m^2_\phi=0}$ plane. Note that, according to (\ref{eq:zeta_uv}) and (\ref{eq:911}), this also corresponds to ${m^2_\phi=0}$.

As one decreases $\mu$ from the chosen initial condition on H1, $\zeta$ decreases while $\tilde m^2$ and $\tilde m^2_\phi$ increase. This means that $\zeta$ converges and that $\tilde m^2$ and/or $\tilde m^2_\phi$ either converge to a finite, non-zero value or diverge. In any case, $\beta^{\rm IR}_\zeta\simeq C_{\rm IR} \zeta^2$ with $C_{\rm IR}>0$ (since the flow occurs in region ${\cal R}_{\uparrow\downarrow}$) and then
\begin{equation}
\zeta(\mu)=\frac{\zeta_0}{1-C_{\rm IR}\zeta_0 \ln\left(\frac{\mu}{\mu_0}\right)}\,,\label{eq:zeta_ir}
\end{equation}
which approaches $0^+$ as ${\mu\to 0}$. It follows that $\beta_{\tilde m^2}$ is dominated by $-2\tilde m^2$ in the IR using the same reasoning as as before. But then (\ref{eq:log}) needs to hold in the IR, which is only possible if ${\tilde m^2\to \infty}$. From a similar argument, we find that ${\tilde m^2_\phi\to \infty}$ and, thus, the beta functions for $\zeta$ and $m^2$ become those in Eq.~\eqref{eq:betairinf}, fixing ${C_{\rm IR}=1/3}$ and implying (\ref{eq:71})-(\ref{eq:72}). 

As for $m^2_\phi$, $\beta_{m^2_\phi}$ approaches $0$ but, because the way it approaches $0$ depends on how fast $\tilde m^2$ and $\tilde m^2_\phi$ diverge with respect to each other, which we do not know at this stage, we need to distinguish two cases as ${\tilde m^2\to\infty}$ and ${\tilde m^2_\phi\to\infty}$. If ${\tilde m^2_\phi\gg\tilde m^2\gg 1}$, we find, see Eq.~(\ref{eq:limits3}),
\begin{equation}
\beta_{m^2_\phi}\sim -4\zeta\frac{m^2_\phi}{\tilde m^2_\phi}\,,
\end{equation}
which rewrites
\begin{equation}
\frac{dm^2_\phi}{d\mu}\sim -4\zeta\mu\,,
\end{equation}
and thus $m_\phi^2\sim -(\zeta/8)\mu^2$ which is inconsistent since $m_\phi^2>0$. We thus need to assume that ${\tilde m^2\gg\tilde m^2_\phi\gg 1}$.\footnote{Strictly speaking, we should also consider the case where $\tilde m^2$ and $\tilde m^2_\phi$ grow at the same rate.} This leads to, see Eq.~(\ref{eq:limits3}),
\begin{equation}
\beta_{m^2_\phi}\sim -2\zeta  m^2_\phi\frac{\ln\tilde m^2}{\tilde m^2}\,,
\end{equation}
which rewrites
\begin{equation}
\frac{d\ln m^2_\phi}{d\ln\mu}\sim -2\zeta \frac{\ln\tilde m^2}{\tilde m^2}\,.
\end{equation}
But now, using that $d\tilde m^2/d\ln\mu$ is dominated by $-2\tilde m^2$ in the IR, we find that
\begin{equation}
\int d\ln\mu \,\zeta \frac{\ln\tilde m^2}{\tilde m^2}~-\frac{1}{2}\int d\tilde m^2 \,\zeta \frac{\ln\tilde m^2}{\tilde m^4}
\end{equation}
converges in the infrared. From a general result on the integration of asymptotic relations, see App.~\ref{app:asymp}, it follows that $\ln m^2_\phi$ admits a limit in the IR and thus that $m^2_\phi$ approaches a non-zero value. 

\begin{figure}[t!]
		\begin{minipage}[b]{0.95\linewidth}
			\includegraphics[width=\textwidth]{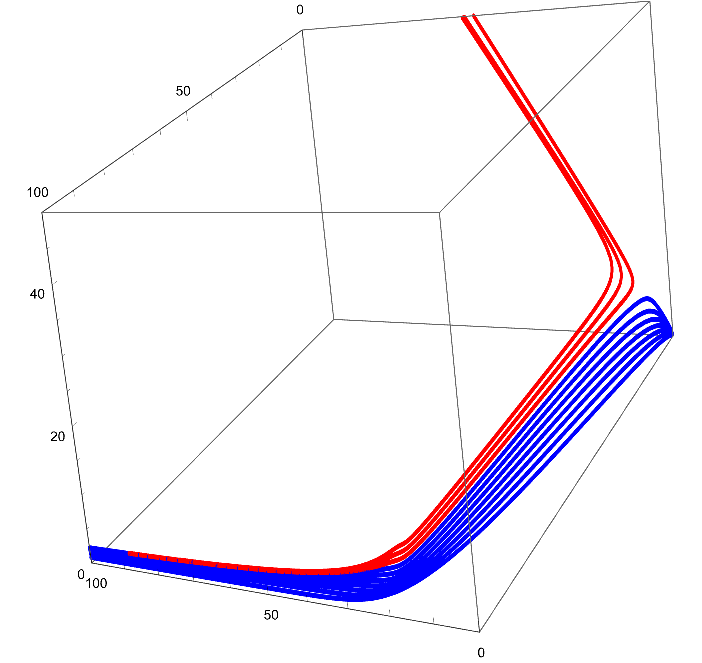}
		\end{minipage}
		\caption{Examples of UV/IR-safe trajectories (blue) and non-UV-safe trajectories (red) in the parameter space $(\tilde m^2,100{\rm Tanh}\,\tilde m^2_\phi,\zeta)$. The former have been obtained by choosing initial points on H1 with ${\tilde m^2_\phi=1}$ and ${\zeta=1}$, $2$, \dots, $7$, while the latter corresponds to a similar initialization with ${\zeta=8}$, $9$ and $10$.}
		\label{1st_look}
\end{figure}

We have thus found that the flow in the bulk of the three-parameter space that starts from the type of considered initial conditions,\footnote{That is, on ${\cal R}_{\uparrow\downarrow}\cap{\cal R}_{\downarrow\downarrow}\cap\{(\tilde m^2,\tilde m^2_\phi,\zeta)\,|\,0 <\zeta \leq 3\}$.}  connects the asymptotically free UV fixed point of the CF model coupled to a massless scalar field (corresponding to $\tilde m^2_\phi=0$ or equivalently $m^2_\phi=0$), to the asymptotically free IR fixed point of the pure CF model (corresponding to $\tilde m^2_\phi=\infty$ but with $m^2_\phi$ finite), see Fig.~\ref{1st_look}.  Once again, it is important to keep in mind that upon including higher-order corrections in this analysis, the RG trajectories studied here could be drastically changed by the development of a possible UV Landau pole. Note that, in the IR, $\zeta$ and $m^2$ run to zero but in the same way. This ensures that the gluon propagator has a decoupling behavior and that the ghost dressing function has a finite limit. Since the scalar mass  approaches a constant, the scalar propagator approaches exactly $1/m^2_{\phi}(\mu_0)$, which we have confirmed numerically.

\section{Comparing with lattice data} \label{SecLattice}

Let us now compare our analytical expressions with available lattice data.
For the scalar propagator, we compare our findings with the lattice data reported in Ref.~\cite{Maas19a}, corresponding to SU(2) YM theory coupled to adjoint scalar matter in four dimensions, for different values of the renormalized lattice scalar mass $m_\phi^{\rm lt}(\mu_0)=x\,{\rm GeV}$, with $x= 0, 0.1, 1$, or $10$, and $\mu_0 = 1.5 \, {\rm GeV}$. 

\begin{figure}[t!]
		\begin{minipage}[b]{1.0\linewidth}
			\includegraphics[width=\textwidth]{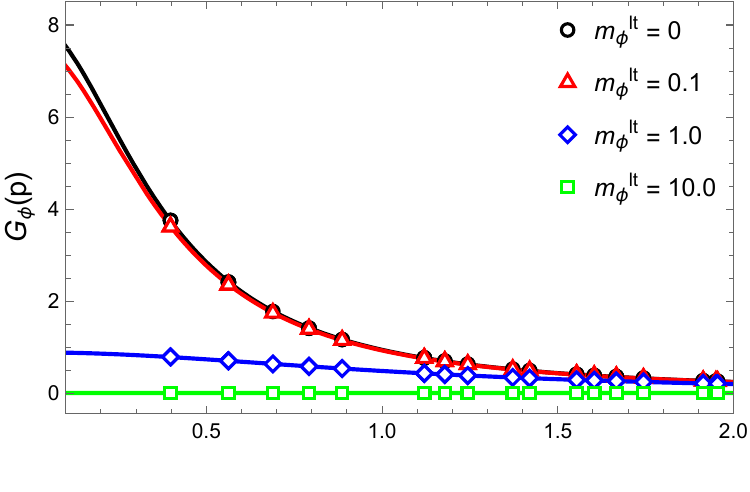}
		\end{minipage}  \hfill
		\caption{Comparison of our analytical results for the scalar propagator with data for the different lattice scalar masses $m_\phi^{\rm lt} (\mu_0)=x\,{\rm GeV}$, with $x=0, 0.1, 1$ or$10$, and $\mu_0=1.5\,{\rm GeV}$.}
		\label{FitAllScalars}
\end{figure}

If we consider only the scalar propagator, we can easily find a very good agreement between our analytical results and the lattice data for the different scalar masses considered, as one can readily see in Fig.~\ref{FitAllScalars}. Indeed, fixing the coupling and gluon mass at the renormalization scale $\mu_0$ as $g_0 = 1.0$ and $m_0 = 0.3 \, {\rm GeV}$ for definiteness, we find remarkably good fits for the four sets of data mentioned above if we choose $m_{\phi0} (\mu_0)=y\,{\rm GeV}$, with $y= 0.36, 0.37, 1.09$ and $10.16$, respectively.  We note that the values of the scalar mass on the lattice and in our continuum approach do not need to coincide since they correspond to renormalized massed in different schemes\footnote{Although we are treating the $m^{\rm lt}=0$ data on the same footing as the other cases in the fitting procedure, we remark that there can be subtleties due to the RG stability of the massless plane.}.

Interestingly enough, at our level of approximation, the momentum dependence of the scalar propagator comes from diagrams involving the gauge coupling and we do not need large values of the latter to reproduce the data remarkably well. However, obtaining good results for the scalar propagator alone was already expected, not representing a great achievement of the CF model, since the standard YM action can already do it at one-loop order without RG improvement, as it was shown in Ref.~\cite{Brito23}. 

The challenge is to find simultaneously good fits for the scalar, gluon and ghost propagators. Unfortunately, to our knowledge, there are no lattice simulations for the gluon and ghost propagators in YM theories coupled with scalar matter. However, as we will argue in the end of this section, we expect a very low sensitivity of gluon and ghost propagators to the presence of scalar matter at leading order. Thus, it is reasonable to use the lattice data for the gluon and ghost propagators without scalar fields as a benchmark to perform the combined fitting procedure. The lattice data for the $SU(2)$ gluon propagator and ghost dressing function in the Landau gauge can be found in~\cite{Cucchieri:2008qm}. Since the corresponding data go only up to $2 \, {\rm GeV}$, we will correspondingly restrict the lattice data for the scalar propagator to this range. 

To compare our results with lattice data, we have to fix the initial conditions of the RG flow which play the role of fitting parameters. They are chosen to minimize simultaneously suitable error functions defined for the scalar, gluon and ghost propagators. The error function associated with a quantity of interest $F$ in the comparison between the analytical and lattice results that we will adopt here is an average between relative and (normalized) absolute errors. It is defined by:
\begin{align}\label{ErrorFunction}
    \chi^2_F = \frac{1}{2N_{\rm lt}} \sum_i \left[F_{\rm th}(p_i) - F_{\rm lt}(p_i)\right]^2 \left[ \frac{1}{F_{\rm lt}(p_i)^{2}} + \frac{1}{F_{\rm lt}(\mu_0)^{2}}\right],
\end{align}
where $N_{\rm lt}$ is the number of lattice points, $F_{\rm th}$ and $F_{\rm lt}$ refer to analytical and lattice values, respectively, and $\mu_0$ is the renormalization point, here used as a reference scale. We also need to include an overall normalization factor that takes into account that the renormalization scheme used in the lattice simulation is different from the one adopted here. This global factor $\mathcal{N}_F$ can be obtained from the minimization of the error function $\chi^2_F$~\eqref{ErrorFunction} defined above, and can be written as:
\begin{align} \label{GlobalFactor}
    \mathcal{N}_F = \frac{\sum_i F_{\rm th}(p_i) F_{\rm lt}(p_i) \left[F_{\rm lt}(p_i)^{-2} + F_{\rm lt}(\mu_0)^{-2}\right]}{\sum_j F_{\rm th}(p_j)^2 \left[F_{\rm lt}(p_j)^{-2} + F_{\rm lt}(\mu_0)^{-2}\right]}.
\end{align}

\vglue2mm

\begin{figure}[t!]
		\begin{minipage}[b]{1.0\linewidth}
			\includegraphics[width=\textwidth]{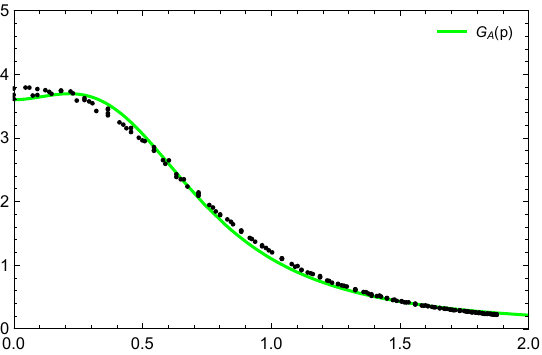}
            \includegraphics[width=\textwidth]{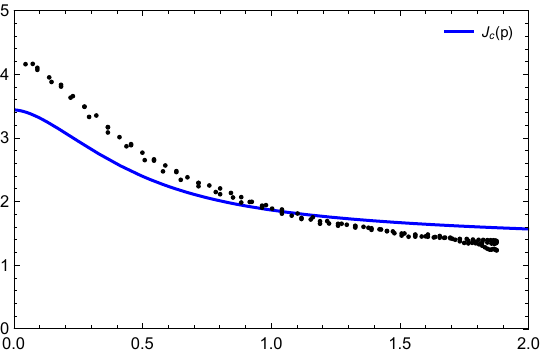}
            \includegraphics[width=\textwidth]{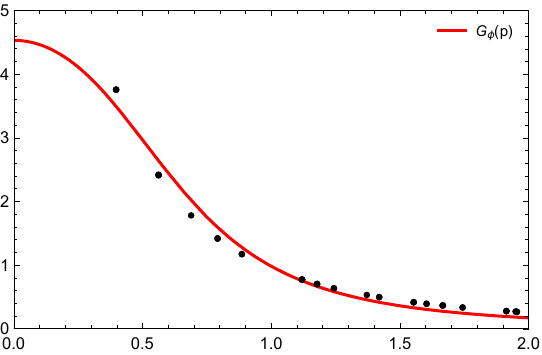}
		\end{minipage}  \hfill
		\caption{CF results for the gluon propagator (top), ghost dressing (middle), and scalar propagator (bottom) compared with lattice data for $m_\phi^{\rm lt} = 0 \, {\rm GeV}$.}
		\label{CompLattMphi0}
	\end{figure}

\begin{figure}[t!]
		\begin{minipage}[b]{1.0\linewidth}
			\includegraphics[width=\textwidth]{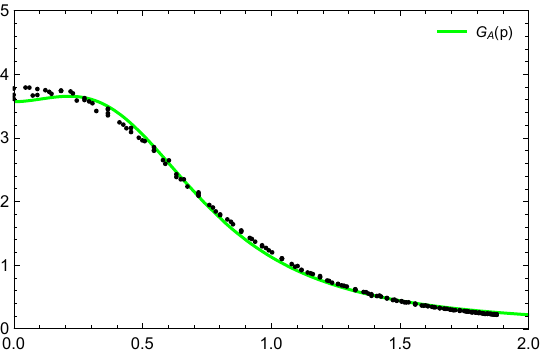}
            \includegraphics[width=\textwidth]{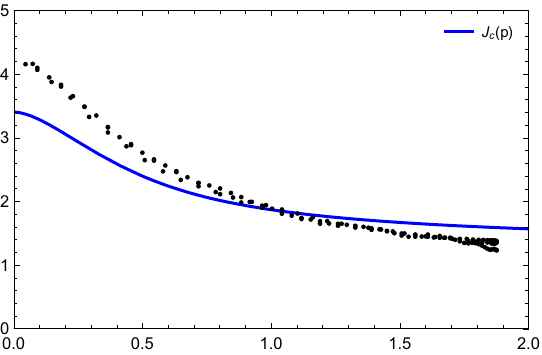}
            \includegraphics[width=\textwidth]{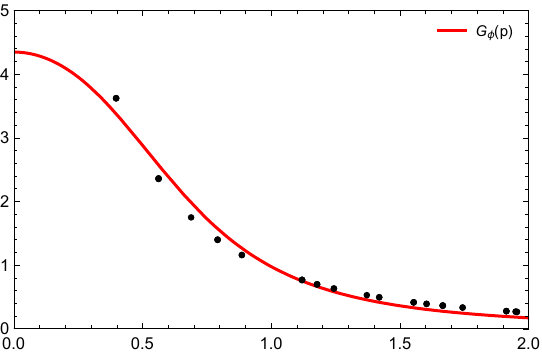}
		\end{minipage}  \hfill
		\caption{CF results for the gluon propagator (top), ghost dressing (middle), and scalar propagator (bottom) compared with lattice data for $m_\phi^{\rm lt} = 0.1 \, {\rm GeV}$.}
		\label{CompLattMphi01}
	\end{figure}

\begin{figure}[t!]
		\begin{minipage}[b]{1.0\linewidth}
			\includegraphics[width=\textwidth]{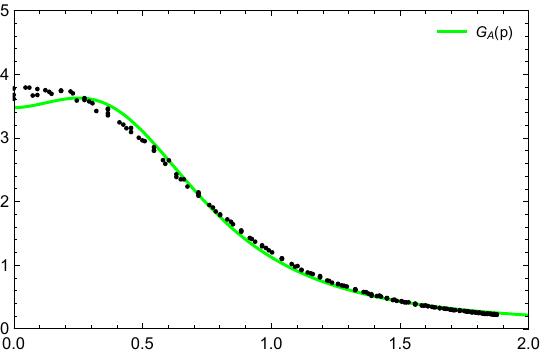}
            \includegraphics[width=\textwidth]{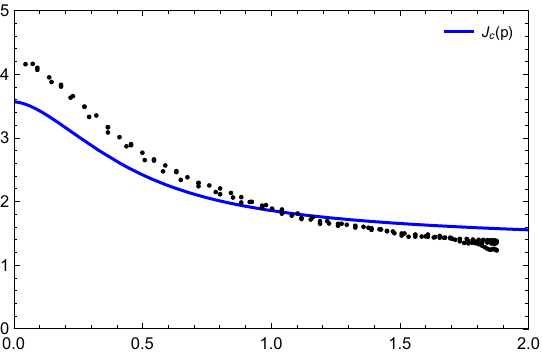}
            \includegraphics[width=\textwidth]{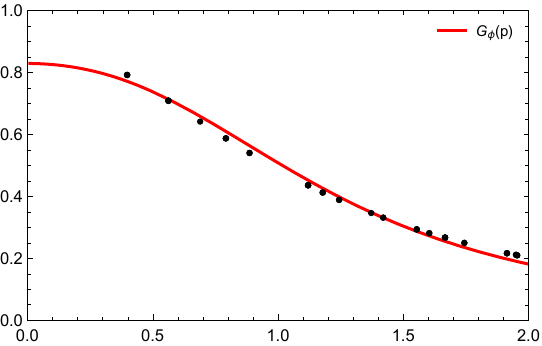}
		\end{minipage}  \hfill
		\caption{CF results for the gluon propagator (top), ghost dressing (middle), and scalar propagator (bottom) compared with lattice data for $m_\phi^{\rm lt} = 1 \, {\rm GeV}$.}
		\label{CompLattMphi1}
	\end{figure}

\begin{figure}[t!]
		\begin{minipage}[b]{1.0\linewidth}
			\includegraphics[width=\textwidth]{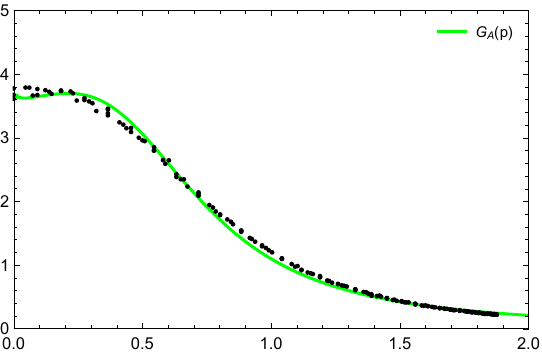}
            \includegraphics[width=\textwidth]{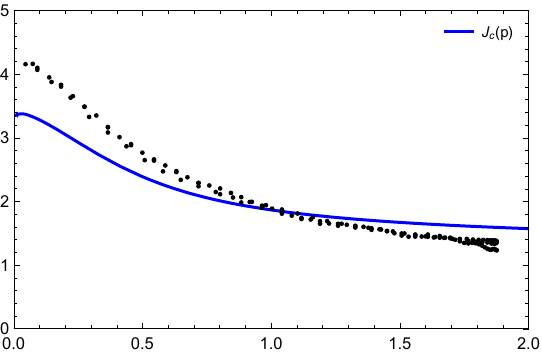}
            \includegraphics[width=\textwidth]{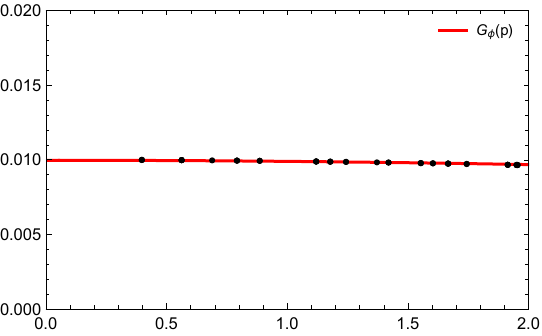}
		\end{minipage}  \hfill
		\caption{CF results for the gluon propagator (top), ghost dressing (middle), and scalar propagator (bottom) compared with lattice data for $m_\phi^{\rm lt} = 10 \, {\rm GeV}$.}
		\label{CompLattMphi10}
	\end{figure}

The routine used for the fitting process was as follows. We start choosing initial values $\left(g_0, m_0, m_{\phi0}\right)$, taken at the renormalization scale $\mu_0 = 1.5 \, {\rm GeV}$, and numerically solve the set of differential equations $da/d\ln\mu= \beta_a$ (for $a = g, m^2, m_\phi^2$), obtaining the RG flow of these parameters. Then, we substitute the parameters $g$, $m^2$ and $m_\phi^2$ in the expressions for $G_A$, $J_c$ and $G_\phi$ by their running counterparts and compute the global factors $\mathcal{N}_A$,  $\mathcal{N}_c$ and $\mathcal{N}_\phi\}$ using Eq.~\eqref{GlobalFactor}. Thus, we compute the associated errors $\chi^2_A$, $\chi^2_c$ and $\chi^2_\phi$ using Eq.~\eqref{ErrorFunction} and take their average $\chi^2_{\rm av}=(\chi^2_A + \chi^2_c + \chi^2_\phi)/3$. We repeat this process for many different initial values $\left(g_0, m_0, m_{\phi0}\right)$ and take the combination which minimizes the averaged error $\chi_{\rm av}$. We follow this procedure for each of the four lattice data sets for the scalar propagator. The best-fit values for each lattice scalar mass is gathered in Table~\ref{BestFitTable}.
The comparison between the CF results at one-loop and the lattice data are shown in Figs.~\ref{CompLattMphi0}, \ref{CompLattMphi01}, \ref{CompLattMphi1}, and \ref{CompLattMphi10}.
\begin{table}[h!]
\centering
\begin{tabular}{clclclclclclc}
\hline \hline
$m_\phi^{\rm lt} (\rm GeV)$         &  & $g_0$  &  & $m_0 (\rm GeV)$  &  & $m_{\phi 0} (\rm GeV)$  &  & $\mathcal{N}_A$   &  & $\mathcal{N}_c$   &  & $\mathcal{N}_\phi$ \\ \hline
$0$    &  & $3.84$ &  & $0.37$ &  & $0.44$       &  & $1.02$ &  & $1.67$ &  & $0.88$  \\
$0.1$  &  & $3.87$ &  & $0.38$ &  & $0.45$       &  & $1.05$ &  & $1.67$ &  & $0.88$  \\
$1$  &  & $3.88$ &  & $0.37$ &  & $1.14$       &  & $1.03$ &  & $1.65$ &  & $1.08$  \\
$10$ &  & $3.81$ &  & $0.37$ &  & $11.55$      &  & $1.01$ &  & $1.67$ &  & $1.33$  \\ \hline \hline
\end{tabular}
\caption{Best-fit parameters $\{g_0, m_0, m_{\phi0}\}$ and $\{\mathcal{N}_A, \mathcal{N}_c, \mathcal{N}_\phi\}$ for the lattice masses $m_\phi^{\rm lt} = 0, 0.1, 1, {\rm and} \, 10 \, {\rm GeV}$.}
\label{BestFitTable}
\end{table}


For all the considered cases, we find a very good agreement between the analytical results at one-loop using the CF model and the lattice data. The values obtained for $\left(g_0, m_0\right)$ in the four cases (see Table~\ref{BestFitTable}) are pretty similar to the results attained for the standard CF case (without scalar matter), which are given by $\left(g_0 = 3.85, m_0 = 0.37 \, {\rm GeV}\right)$, showing that these parameters are essentially constrained by the lattice data in the gauge sector. That is, since the gauge sector is more sensitive to $g$ and $m$ than the scalar one, the restriction in their range coming from the former is stricter.   

Let us now we discuss the impact of scalar matter on the gluon propagator and ghost dressing function, aiming to justify that it is a reasonable approximation to use lattice data for these objects obtained in the absence of scalar matter as a benchmark for the combined fit with the scalar propagator at leading order. 

The direct contribution of the scalar fields to the gauge sector is done through the object $\delta \gamma_A = \zeta f_3$, which enters both in $\beta_\zeta$ and in $\beta_{m^2}$. Rewriting $f_3$ in terms of the dimensionful mass $m_\phi^2 = \mu^2 \tilde{m}_\phi^2$ and looking at its expression for small values of $\mu$, one can see that its IR behavior is of the form $\mu^2 /30 m_\phi^2$. From this, one concludes  that, in the IR limit, the scalar contribution to the gauge sector decreases as fast as $\mu^2$ when $\mu \to 0$. Moreover, for a fixed value of $\mu$ it gets smaller as we increase the value of $m_\phi$.

To numerically support this conclusion, one can explicitly compare the results for the gluon propagator and the ghost dressing function computed at one-loop order with the CF model in the scenarios with and without scalar matter. To do so, we introduce the relative difference between the quantities of interest, defined as
\begin{align} \label{RelativDiffPureCF}
    \Delta G_A (p) &= \frac{G_A(p) - G_A^{\#}(p)}{G_A^{\#}(p)} \nonumber \\
    \Delta J_c (p) &= \frac{J_c(p) - J_c^{\#}(p)}{J_c^{\#}(p)},
\end{align}
where $G_A^{\#}$ and $J_c^{\#}$ are the gluon propagator and ghost dressing obtained from the pure CF model, without scalar matter. For the initial values considered here, the relative difference remains below five percent, which is totally acceptable for a one-loop computation. It is also worth pointing out that the discrepancies $\Delta G_A$ and $\Delta J_c$ get smaller as we increase $m_{\phi 0}$. This is intuitively expected since as we increase the scalar mass, we are somehow decoupling the scalar field from the IR physics. 

\begin{figure}[t!]
		\begin{minipage}[b]{1.0\linewidth}
			\includegraphics[width=\textwidth]{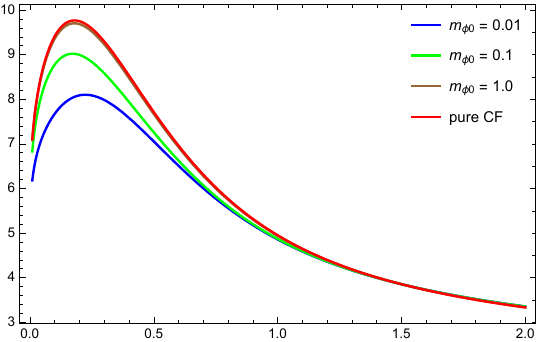}
            \includegraphics[width=\textwidth]{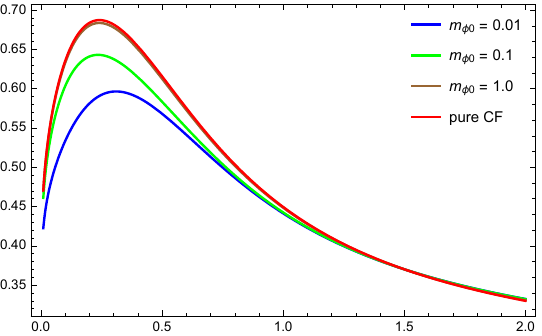}
		\end{minipage}  \hfill
		\caption{Comparison between the RG flows of $g(\mu)$ (top) and $m(\mu)$ (bottom) with initial values $g_0 = 3.85$ and $m_0 = 0.37$, for $m_{\phi 0} = 0.01, 0.1, \, {\rm and} \, 1 \, {\rm GeV}$ together with the pure CF model (without scalar matter).}
		\label{RGFlowsDiffMphi}
\end{figure}

The behavior described above can be understood in a simple manner. As one can readily see in Eq.~\eqref{FinalPropagRG}, the gluon and ghost two-point function behaviors are determined by the coupling and mass RG flows $g(\mu)$ and $m(\mu)$. Fixing the initial values $\left(g_0 = 3.85, m_0 = 0.37\,  {\rm GeV}\right)$ for definiteness, we can consider different values for $m_{\phi 0}$ to see what is its impact in the RG flows $g(\mu)$ and $m(\mu)$. In Fig.~\ref{RGFlowsDiffMphi} we exhibit the RG flows $g(\mu)$ and $m(\mu)$ for $m_{\phi 0} = 0.01, 0.1,\, {\rm and} \, 1 \, {\rm GeV}$ and compare these results with the corresponding RG flows obtained for the pure CF model for the same initial values. As one can read from the figures, as we increase the value of $m_{\phi 0}$, the RG flows $g(\mu)$ and $m(\mu)$ converge to the one obtained without scalar matter, confirming the described behavior.   

\section{Comments on the RGZ results}
\label{SecRGZ}

In this section, we briefly comment on the results obtained using the RGZ framework, contrasting them with our findings in the CF model. As discussed in Ref.~\cite{Brito23}, two rather different strategies have been considered in the literature for coupling colored matter fields to the RGZ action: a minimal coupling, in which the ordinary matter action is kept untouched and the restriction to the Gribov region affects the matter sector only through loop corrections~\cite{Brito23}; a non-minimal coupling, in which a matter horizon-like function is introduced in direct analogy with the gauge horizon function~\cite{Capri14, Capri17}. We focus mostly on the former, since it is the strategy directly comparable to the one adopted in this work, and comment more briefly on the latter at the end of this section.

The minimal-coupling strategy is the one pursued in Ref.~\cite{Brito23}. There, the tree-level scalar propagator is left untouched, and the Gribov horizon enters the scalar two-point function only through the one-loop diagrams built with the RGZ gluon propagator. The comparison with that work is particularly natural: both analyses address the IR regime of YM theories minimally coupled to scalar matter in the adjoint representation, compute the scalar propagator at one loop, and confront the result with the same lattice data~\cite{Maas19a}, for the same set of scalar masses.

The two analyses differ, however, in an essential way. In the RGZ framework, the mass parameters entering the gluon propagator, $(\gamma,m^2,M^2)$, are fixed once and for all by an independent fit of that tree-level propagator to lattice data and are then simply inserted into the one-loop scalar propagator.  It is worth mentioning that, currently, no RG improvement analogous to the one implemented here is available in that framework. As a consequence, the comparison with the scalar lattice data in Ref.~\cite{Brito23} is carried out using the scalar sector alone, without a simultaneous one-loop fit of the gluon and ghost propagators. 

In the present work, by contrast, the combination of the IR-safe renormalization scheme with the RG flow allows for a genuinely simultaneous one-loop description of the gluon, ghost, and scalar propagators, without the need of any artificial choice of parameters to match the data. However, it is important to keep in mind that the lattice data for gluon and ghost propagators were simulated in the absence of scalar matter. 

Let us close with a few comments on the non-minimal coupling proposed in Refs.~\cite{Capri14, Capri17}. There, one introduces a horizon-like function for the matter field, built in analogy with the gauge horizon function, together with an additional parameters and auxiliary fields. The resulting tree-level scalar propagator is already capable of reproducing the qualitative features seen in the lattice data even without considering one-loop corrections. The drawback of this approach is that one has to introduce a set of auxiliary fields and mass parameters which must be fixed by hand, without a clear justification for it.

\section{Concluding remarks} \label{SecConclusion}

In this work, we provided a first exploration of YM theories coupled to scalar matter within the CF framework. Our main goal was to investigate whether this remarkably successful perturbative approach could be extended to the case with scalar matter, assessing its ability to describe the IR dynamics of such a system efficiently.

We presented the CF model coupled to scalar matter and its main properties. We computed the two-point correlation functions at one-loop and discussed the renormalization of this model in two schemes of interest: i) the minimal subtraction scheme; ii) a generalization of the infrared-safe scheme usually adopted in the standard CF model, which is an important ingredient for its success.

Particular attention was devoted to the RG structure of the theory. The resulting RG analysis revealed a rich IR structure. In the limit of vanishing scalar mass, we uncovered novel RG fixed points; for the limit of infinite scalar mass, we recovered the standard CF model results; when considering finite non-vanishing scalar masses, our analysis indicates the absence of fixed-points. Remarkably, we found the existence of infrared-safe trajectories, implying that it is possible to have a running coupling that remains finite and under perturbative control. This provides strong evidence that perturbation theory around a massive extension of the FP action remains a viable tool even in the presence of scalar matter. 

Another central result of this work is the successful comparison between our one-loop analytical results and non-perturbative information obtained through lattice simulations. Indeed, we performed a simultaneous fit for the scalar, gluon, and ghost propagators, finding an excellent agreement with lattice data. Although lattice results for gluon and ghost propagators in YM theories coupled to adjoint scalars are presently unavailable, we argued that the contributions due to the existence of scalar matter to these correlators are sufficiently small to remain below the typical uncertainty associated with a one-loop calculation, making the comparison meaningful within the accuracy of our approximation.

However, as already stressed in the introduction, the results presented here must be taken with a grain of salt. In fact, in our one-loop approximation, the propagators are independent of the scalar self-interaction. Thus, our results are insensitive to one dimension of the parameter space and could be drastically changed depending on the behavior of the running scalar coupling, which we plan to investigate in the future.

The present work should be viewed only as a first exploration of YM theories coupled with scalar matter from the CF model perspective. The success of the framework in simultaneously describing the lattice data for the scalar, gluon, and ghost propagators, together with the emergence of infrared-safe RG trajectories, provides further support for the idea that it is possible to use perturbation theory reliably to extract information from the IR regime. In this sense, this work opens a new avenue for investigating the CF framework in more general scenarios, offering encouraging evidence of its versatility as an effective description of strongly interacting YM theories.

It would be highly desirable to consider higher-order corrections for the propagators within this approach. In this way, the scalar self-coupling will contribute to the propagators, enabling us to consider all RG equations together and explore the RG flows in the full parameter space of this theory. One could wonder if there exists some strategy within the CF framework that would enable us to handle the famous Landau pole present in the scalar self-coupling, inspired by its success in avoiding the well-known Landau pole in the gauge coupling that jeopardizes perturbation theory using the FP action. In this respect, we stress the importance of exploring different renormalization schemes. Indeed, some regions of the parameter space might only be reachable within particular schemes. The pure CF model is a good illustration of this fact. Another interesting path to follow would be to consider the Higgs phase of this system, discussing different breaking patterns. Finally, the finite-temperature version of this system will allow many interesting investigations. These are directions for future work.

\begin{acknowledgments}
    The authors are grateful to Axel Maas and Tereza Mendes for providing the lattice data used in this work. The authors also thank Nicolás Wschebor, Nahuel Barrios, Gustavo P. de Brito, and Antônio D. Pereira for useful discussions. We acknowledge the financial support from the Program for the Development of Basic Sciences (PEDECIBA), the ECOS program, and ANII through the FCE-2025-186497. P.~De~Fabritiis acknowledges the National Council for Scientific and Technological Development - CNPq for the financial support (No. 402459/2024-5). G. L. L. W. Levy acknowledges financial support from CAPES. This study was financed in part by the Coordenação de Aperfeiçoamento de Pessoal de Nível Superior - Brasil (CAPES) - Finance Code 001.
\end{acknowledgments}

\appendix 

\section{Feynman rules}\label{AppFeynman}

\begin{figure}[h!]
\centering
\begin{equation}
\begin{aligned}
\raisebox{-0.35\height}{
\includegraphics[width=0.25\columnwidth]{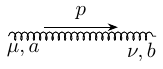}
} = \delta^{ab} \left(\frac{1}{p^2+m^2}\right) P^\perp_{\mu\nu}(p)
\end{aligned}
\end{equation}
\end{figure}
\begin{figure}[h!]
\centering
\begin{equation}
\begin{aligned}
\raisebox{-0.35\height}{
\includegraphics[width=0.25\columnwidth]{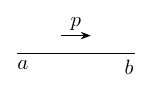}
} = \delta^{ab}\frac{1}{p^2 + m_\phi^2}
\end{aligned}
\end{equation}
\end{figure}
\begin{figure}[h!]
\centering
\begin{equation}
\begin{aligned}
\raisebox{-0.35\height}{
\includegraphics[width=0.25\columnwidth]{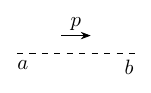}
} = \delta^{ab}\frac{1}{p^2} 
\end{aligned}
\end{equation}
\end{figure}
\begin{figure}[h!]
\centering
\begin{equation}
\begin{aligned}
\raisebox{-0.35\height}{
\includegraphics[width=0.25\columnwidth]{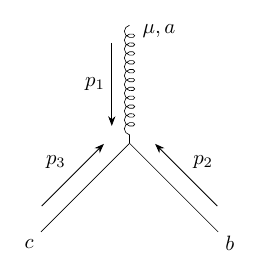}
} = i g f^{abc} (p_2 - p_3)^\mu
\end{aligned}
\end{equation}
\end{figure}
\begin{figure}[h!]
\centering
\begin{equation}
\begin{aligned}
\raisebox{-0.35\height}{
\includegraphics[width=0.25\columnwidth]{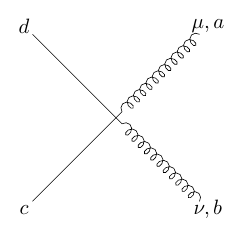}
} = g^2 \delta_{\mu \nu} (f^{adx}f^{bcx}+ f^{acx}f^{bdx})
\end{aligned}
\end{equation}
\end{figure}
\begin{figure}[h!]
\centering
\begin{equation}
\begin{aligned}
\raisebox{-0.35\height}{
\includegraphics[width=0.2\columnwidth]{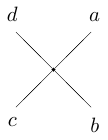}
} = \frac{\lambda}{3}\biggl(\delta^{ab}\delta^{cd}+ \delta^{ac}\delta^{bd}+\delta^{ad}\delta^{bc}\biggl)
\end{aligned}
\end{equation}
\end{figure}
\begin{figure}[h!]
\centering
\begin{equation}
\begin{aligned}
\raisebox{-0.35\height}{
\includegraphics[width=0.25\columnwidth]{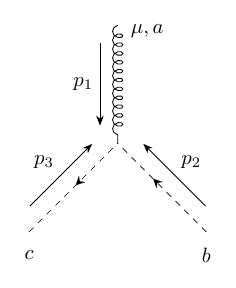}
} = -i g f^{abc} p_3^\mu
\end{aligned}
\end{equation}
\end{figure}
\begin{figure}[h!]
\centering
\begin{equation}
\begin{aligned}
 \raisebox{-0.35\height}{
\includegraphics[width=0.3\columnwidth]{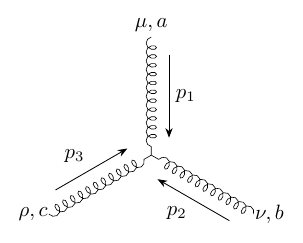}
}&= igf^{abc}\left[\delta_{\mu \nu} (p_1-p_2)^\rho + \right.\\ 
        &\left. \delta_{\mu \rho} (p_3-p_1)^{\nu} + \delta_{\nu \rho}(p_2 -p_3)^{\mu}\right]
\end{aligned}
\end{equation}
\end{figure}
\begin{figure}[h!]
\centering
\begin{equation}
\begin{aligned}
\raisebox{-0.35\height}{
\includegraphics[width=0.3\columnwidth]{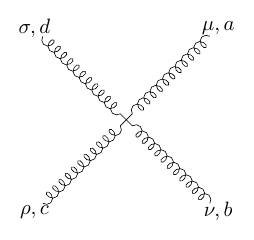}
} &= g^2\big[f^{adx}f^{bcx}(\delta_{\mu \nu}\delta_{\rho \sigma}-\delta_{\mu \rho}\delta_{\nu \sigma})+\\
         &f^{acx}f^{bdx}(\delta_{\mu \nu}\delta_{\rho \sigma}-\delta_{\mu \sigma}\delta_{\nu \rho})+  \\
      &f^{abx}f^{cdx}(\delta_{\mu \rho}\delta_{\nu \sigma}-\delta_{\mu \sigma}\delta_{\nu \rho})\big]
\end{aligned}
\end{equation}
\end{figure}

\pagebreak

\section{Asymptotic behavior of the flow}\label{app:asymp}

Let us here derive more carefully the various asymptotic behaviors of the RG flow in the present model. To this purpose, we shall use the following well know result. Consider two functions $f(x)$ and $g(x)$ such that ${f(x)\sim g(x)}$ as ${x\to x_0}$. If $\int^x dy\,f(y)$ diverges as ${x\to x_0}$ then so does $\int^x dy\,g(y)$, and
\begin{equation}
\int^x dy\,f(y)\sim \int^x dy\,g(y)\,.
\end{equation}
Similarly, if $\int^x dy\,f(y)$ converges as ${x\to x_0}$ then so does $\int^x dy\,g(y)$, and
\begin{equation}
\int^{x_0}_x dy\,f(y)\sim \int^{x_0}_x dy\,g(y)\,.
\end{equation}
We will also need to exponentiate asymptotic relations such as ${f(x)\sim g(x)}$. Here, it should be stressed that it is not necessarily true that ${\exp(f(x))\sim\exp(g(x))}$. In fact, this later relation holds true iff $f(x)-g(x)\to 0$ which it is mandatory to verify before exponentiating ${f(x)\sim g(x)}$.\\

Consider first the boundary ${\tilde m^2_\phi=\infty}$ corresponding to the pure CF model. The flow equations write
\begin{eqnarray}
\beta_\zeta & = & \zeta^2 F(\tilde m^2)\,,\\
\beta_{\tilde m^2} & = & \tilde m^2(\zeta G(\tilde m^2)-2)\,,
\end{eqnarray}
with ${F\equiv f_1+2f_2}$ and ${G=f_1+f_2}$. Let us focus on the family of IR-safe trajectories which are such that ${\tilde m^2\to 0}$ in the UV and ${\tilde m^2\to \infty}$ in the IR, while ${\zeta\to 0^+}$ both in the UV and in the IR. We also recall that $F(0)$ and $G(0)$ are both strictly negative while $F(\infty)$ and $G(\infty)$ are both strictly positive.

In the UV, since ${\tilde m^2\to 0}$ and ${\zeta\to 0^+}$, we can write at leading order
\begin{eqnarray}
\beta_\zeta & \sim & \zeta^2 F(0)\,,\\
\beta_{\tilde m^2} & \sim & -2\tilde m^2\,.
\end{eqnarray}
The first equation rewrites
\begin{equation}
\frac{d(1/\zeta)}{d\ln\mu}\sim -F(0)\,.\label{eq:logdim}
\end{equation}
Since the $\ln\mu$ integral of the right-hand side diverges in the UV, 
it follows from the general result recalled above that\footnote{We are treating the asymptotic expansion in a mathematical sense, without paying attention to physical dimensions. Of course, from a physical point of view the argument of the logarithm should carry no dimension. This will be the result of the next order in the asymptotic expansion which provides the constant under the logarithm, see below.}
\begin{equation}
\frac{1}{\zeta}\sim -F(0)\ln\mu\,,\label{eq:asymp1}
\end{equation}
which is indeed compatible with ${\zeta\to 0^+}$ in the UV. 

From the second flow equation, we obtain similarly
\begin{equation}
\frac{d\ln\tilde m^2}{d\ln\mu}\to -2\,,\label{eq:mt}
\end{equation}
and thus
\begin{equation}
\ln\tilde m^2\sim -2\ln\mu\,,\label{eq:asymp2}
\end{equation}
which is again compatible with $\tilde m^2\to 0$ in the UV. 

It is now tempting, but not necessarily correct, to exponentiate the asymptotic estimates (\ref{eq:asymp1}) and (\ref{eq:asymp2}) to deduce $\mu$ as a function of $\zeta$ in the UV or to conclude that $\tilde m^2$ scales like $1/\mu^2$ and thus that $m^2$ approaches a constant. As matter of fact this second conclusion is erroneous and can be traced back to the fact that asymptotic relations can only be exponentiated in specific situations as recalled above. For instance, in order to exponentiate (\ref{eq:asymp1}), one need to make sure that the subleading term goes to $0$ in the UV. To investigate this, we consider the beta function at next-to-leading order\footnote{The ${\cal O}(\tilde m^4)$ implicitly include possible logarithms of $\tilde m^2$.}
\begin{equation}
\beta_\zeta=\zeta^2(F(0)+(A+B\ln \tilde m^2)\tilde m^2+ {\cal O}(\tilde m^4))
\end{equation}
which we rewrite as
\begin{equation}
\frac{d(1/\zeta+F(0)\ln\mu)}{d\ln\mu}\sim -(A+B\ln \tilde m^2)\tilde m^2\,.\label{eq:nlo}
\end{equation}
We note that
\begin{eqnarray}
& & \int d\ln\mu\,(A+B\ln \tilde m^2)\tilde m^2\nonumber\\
& & \hspace{0.5cm}=\int d\tilde m^2 \frac{d\ln\mu}{d\tilde m^2}(A+B\ln \tilde m^2)\tilde m^2\nonumber\\
& & \hspace{0.5cm}=\,-\frac{1}{2}\int d\tilde m^2 (A+B\ln \tilde m^2)\,,
\end{eqnarray}
where we used the leading order expression for $\beta_{\tilde m^2}$. It follows that the $\ln\mu$ integral of the right-hand side of (\ref{eq:nlo}) converges in the UV. From the result recalled above, it follows that $1/\zeta+F(0)\ln\mu$ converges in the UV and that
\begin{equation}
\left[\frac{1}{\zeta}+F(0)\ln\mu\right]^\infty_\mu\sim\,\frac{1}{2}[(A-B+B\ln\tilde m^2)\tilde m^2]^\infty_\mu\,.
\end{equation}
Without loss of generality, we can write
\begin{equation}
\lim_{\mu\to\infty}\left[\frac{1}{\zeta}+F(0)\ln\mu\right]\equiv F(0)\ln \Lambda\,.
\end{equation}
Using that
\begin{equation}
\lim_{\mu\to\infty}\tilde m^2=0\,,
\end{equation}
we arrive at
\begin{equation}
\frac{1}{\zeta}+F(0)\ln\frac{\mu}{\Lambda}\sim \frac{1}{2}(A-B+B\ln\tilde m^2)\tilde m^2\,.
\end{equation}
This fixes the missing constant under the logarithm in (\ref{eq:logdim}) and shows that the asymptotic relation can be exponentiated into
\begin{equation}
\mu\sim\Lambda e^{-\frac{1}{F(0)}\frac{1}{\zeta}}\,.
\end{equation}
As for $\tilde m^2$, in order to show that the exponentiation of (\ref{eq:mt}) leads to a wrong result, let us consider the beta function for $m^2$ at leading order:
\begin{equation}
\beta_{m^2}\sim\zeta m^2 G(0)\,,
\end{equation}
which rewrites
\begin{equation}
\frac{d\ln m^2}{d\ln\mu}\sim -\frac{G(0)}{F(0)}\frac{1}{\ln\frac{\mu}{\Lambda}}\,.
\end{equation}
Since the $\ln\mu$-integral of the right-hand side diverges, we deduce that
\begin{equation}
\ln m^2\sim \ln\left(\ln\frac{\mu}{\Lambda}\right)^{-\frac{G(0)}{F(0)}}\,.\label{eq:iii}
\end{equation}
Without further information, we are not allowed to exponentiate this relation but we see already that $m\to 0$ in the UV, unlike what the naive exponentiation of (\ref{eq:asymp2}) would lead us to deduce.

To see whether we can exponentiate (\ref{eq:iii}), we need to consider $\beta_{m^2}$ at next-to-leading order
\begin{eqnarray}
\beta_{m^2} & = & \zeta m^2 (G(0)+(C+D\ln\tilde m^2)\tilde m^2+{\cal O}(\tilde m^4))\nonumber\\\nonumber\\
& = & m^2\frac{G(0)+(C+D\ln\tilde m^2)\tilde m^2+{\cal O}(\tilde m^4)}{-F(0)\ln\frac{\mu}{\Lambda}+\frac{1}{2}(A-B+B\ln\tilde m^2)\tilde m^2+{\cal O}(\tilde m^4)}\nonumber\\
& = & -\frac{m^2}{\ln\frac{\mu}{\Lambda}}\frac{G(0)}{F(0)}\frac{1+\frac{C+D\ln\tilde m^2}{G(0)}\tilde m^2+{\cal O}(\tilde m^4)}{1-\frac{1}{2}\frac{A-B+B\ln\tilde m^2}{F(0)}\frac{\tilde m^2}{\ln\frac{\mu}{\Lambda}}+{\cal O}(\tilde m^4)}\nonumber\\
& = & -\frac{m^2}{\ln\frac{\mu}{\Lambda}}\frac{G(0)}{F(0)}\left[1+\left(\frac{C+D\ln\tilde m^2}{G(0)}\right.\right.\nonumber\\
& & \hspace{0.5cm}\left.\left.+\,\frac{1}{2}\frac{(A-B+B\ln\tilde m^2)}{F(0)\ln\frac{\mu}{\Lambda}}\right)\tilde m^2+{\cal O}(\tilde m^4)\right],
\end{eqnarray}
which we rewrite as
\begin{equation}
\frac{d}{d\ln\mu}\left(\ln m^2-\ln\left(\ln\frac{\mu}{\Lambda}\right)^{-\frac{G(0)}{F(0)}}\right)\sim -\frac{(C+D\ln\tilde m^2)\tilde m^2}{F(0)\ln\frac{\mu}{\Lambda}}\,.
\end{equation}
Since the right-hand side is integrable in the UV, we deduce that $\ln m^2-\ln (\ln \mu/\Lambda)^{-G(0)/F(0)}$ converges to a quantity that we denote $\ln M^2$ and then
\begin{eqnarray}
\ln\frac{m^2}{M^2} & - & \ln\left(\ln\frac{\mu}{\Lambda}\right)^{-\frac{G(0)}{F(0)}}\nonumber\\
& \sim & \int_{\mu}^\infty d\ln\nu\,\frac{(C+D\ln\tilde m^2(\nu))\,\tilde m^2(\nu)}{F(0)\ln \frac{\nu}{\Lambda}}\,.
\end{eqnarray}
Since the right-hand side approaches $0$, we deduce that
\begin{equation}
m^2\sim M^2\left(\ln\frac{\mu}{\Lambda}\right)^{-\frac{G(0)}{F(0)}}\,.
\end{equation}
A similar analysis can be done in the IR where ${\zeta\to 0^+}$ while ${\tilde m^2\to\infty}$.



\begin{thebibliography}{0}%
\makeatletter
\providecommand \@ifxundefined [1]{%
 \@ifx{#1\undefined}
}%
\providecommand \@ifnum [1]{%
 \ifnum #1\expandafter \@firstoftwo
 \else \expandafter \@secondoftwo
 \fi
}%
\providecommand \@ifx [1]{%
 \ifx #1\expandafter \@firstoftwo
 \else \expandafter \@secondoftwo
 \fi
}%
\providecommand \natexlab [1]{#1}%
\providecommand \enquote  [1]{``#1''}%
\providecommand \bibnamefont  [1]{#1}%
\providecommand \bibfnamefont [1]{#1}%
\providecommand \citenamefont [1]{#1}%
\providecommand \href@noop [0]{\@secondoftwo}%
\providecommand \href [0]{\begingroup \@sanitize@url \@href}%
\providecommand \@href[1]{\@@startlink{#1}\@@href}%
\providecommand \@@href[1]{\endgroup#1\@@endlink}%
\providecommand \@sanitize@url [0]{\catcode `\\12\catcode `\$12\catcode `\&12\catcode `\#12\catcode `\^12\catcode `\_12\catcode `\%12\relax}%
\providecommand \@@startlink[1]{}%
\providecommand \@@endlink[0]{}%
\providecommand \url  [0]{\begingroup\@sanitize@url \@url }%
\providecommand \@url [1]{\endgroup\@href {#1}{\urlprefix }}%
\providecommand \urlprefix  [0]{URL }%
\providecommand \Eprint [0]{\href }%
\providecommand \doibase [0]{https://doi.org/}%
\providecommand \selectlanguage [0]{\@gobble}%
\providecommand \bibinfo  [0]{\@secondoftwo}%
\providecommand \bibfield  [0]{\@secondoftwo}%
\providecommand \translation [1]{[#1]}%
\providecommand \BibitemOpen [0]{}%
\providecommand \bibitemStop [0]{}%
\providecommand \bibitemNoStop [0]{.\EOS\space}%
\providecommand \EOS [0]{\spacefactor3000\relax}%
\providecommand \BibitemShut  [1]{\csname bibitem#1\endcsname}%
\let\auto@bib@innerbib\@empty
\end{thebibliography}%


\begin{thebibliography}{99}

\bibitem{Yang54}
C. N. Yang and R. L. Mills, {\it Conservation of Isotopic Spin and Isotopic Gauge Invariance}, Phys. Rev. 96, 191 (1954).

\bibitem{tHooft07}
G. ’t Hooft, {\it The making of the standard model}, Nature 448, 271 (2007).

\bibitem{Gross23}
F. Gross, E. Klempt, S. J. Brodsky, et al., {\it 50 Years of quantum chromodynamics}, Eur. Phys. J. C 83, 1125 (2023).

\bibitem{Brambilla14}
N. Brambilla, S. Eidelman, P. Foka, et al., {\it QCD and strongly coupled gauge theories: challenges and perspectives}, Eur. Phys. J. C 74, 2981 (2014).

\bibitem{Politzer73}
H. D. Politzer, {\it Reliable perturbative results for strong interactions?}, Phys. Rev. Lett. 30, 1346 (1973).

\bibitem{Gross73}
D. J. Gross and F. Wilczek, {\it Ultraviolet behavior of non-Abelian gauge theories}, Phys. Rev. Lett. 30, 1343 (1973).

\bibitem{Greensite11}
J. Greensite, {\it An Introduction to the Confinement Problem}, Lecture Notes in Physics Vol. 821 (Springer, New York, 2011).


\bibitem{vonSmekal97}
L. von Smekal, R. Alkofer, and A. Hauck, The infrared behavior of gluon and ghost propagators in Landau gauge QCD, Phys. Rev. Lett. 79, 3591 (1997).

\bibitem{Alkofer01}
R. Alkofer and L. von Smekal, The infrared behavior of QCD Green’s functions: Confinement dynamical symmetry breaking, and hadrons as relativistic bound states, Phys. Rep. 353, 281 (2001).

\bibitem{Aguilar08}
A. C. Aguilar, D. Binosi, and J. Papavassiliou, Gluon and ghost propagators in the Landau gauge: Deriving lattice results from Schwinger-Dyson equations, Phys. Rev. D 78, 025010 (2008).

\bibitem{Alkofer10}
R. Alkofer, M. Q. Huber, and K. Schwenzer, Infrared singularities in Landau gauge Yang-Mills theory, Phys. Rev. D 81, 105010 (2010).

\bibitem{Huber20}
M. Q. Huber, Nonperturbative properties of Yang–Mills theories, Phys. Rep. 879, 1 (2020).

\bibitem{Pawlowski04}
J. M. Pawlowski, D. F. Litim, S. Nedelko, and L. von Smekal, Infrared behavior and fixed points in Landau gauge QCD, Phys. Rev. Lett. 93, 152002 (2004).

\bibitem{Fischer09}
C. S. Fischer, A. Maas, and J. M. Pawlowski, On the infrared behavior of Landau gauge Yang-Mills theory, Ann. Phys. (N.Y.) 324, 2408 (2009).

\bibitem{Dupuis21}
N. Dupuis, L. Canet, A. Eichhorn, W. Metzner, J.M. Pawlowski, M. Tissier, and N. Wschebor, The nonperturbative functional renormalization group and its applications, Phys. Rep. 910, 1 (2021).

\bibitem{Cyrol16}
A. K. Cyrol, L. Fister, M. Mitter, J. M. Pawlowski, and N. Strodthoff, Landau gauge Yang-Mills correlation functions, Phys. Rev. D 94, 054005 (2016).

\bibitem{Cyrol18}
A. K. Cyrol, J. M. Pawlowski, A. Rothkopf, and N. Wink, Reconstructing the gluon, SciPost Phys. 5, 065 (2018).



\bibitem{Wilson74}
K. G. Wilson, {\it Confinement of quarks}, Phys. Rev. D 10, 2445 (1974).

\bibitem{Creutz83}
M. Creutz, {\it Quarks, Gluons and Lattices}, (Cambridge University Press, Cambridge, England, 1983).

\bibitem{Bogolubsky09}
I. L. Bogolubsky, E.-M. Ilgenfritz, M. Müller-Preussker, and A. Sternbeck, {\it Lattice gluodynamics computation of Landau gauge Green’s functions in the deep infrared}, Phys. Lett. B 676, 69 (2009).

\bibitem{Faddeev67}
L. Faddeev and V. Popov, {\it Feynman diagrams for the Yang-Mills field}, Phys. Lett. 25B, 29 (1967).

\bibitem{Gribov78}
V. N. Gribov, {\it Quantization of non-Abelian gauge theories}, Nucl. Phys. B139, 1 (1978).

\bibitem{Vandersickel12}
N. Vandersickel and D. Zwanziger, {\it The Gribov problem and QCD dynamics}, Phys. Rept. 520, 175 (2012).

\bibitem{Zwanziger89}
D. Zwanziger, {\it Local and renormalizable action from the Gribov horizon}, Nucl. Phys. B 323, 513 (1989).

\bibitem{DellAntonio91}
G. Dell'Antonio and D. Zwanziger, {\it Every gauge orbit passes inside the Gribov Horizon}, Commun. Math. Phys. 138, 291 (1991).

\bibitem{Singer78}
I. Singer, {\it Some remarks on the Gribov ambiguity}, Commun. Math. Phys. 60, 7 (1978).

\bibitem{Capri13}
M. A. L. Capri, D. Dudal, M. S. Guimaraes, L. F. Palhares, and S. P. Sorella, {\it An all-order proof of the equivalence between Gribov's no-pole and Zwanziger's horizon conditions}, Phys. Lett. B 719, 448 (2013).

\bibitem{Dudal08}
D. Dudal, J. A. Gracey, S. Sorella, N. Vandersickel, and H. Verschelde, {\it Refinement of the Gribov-Zwanziger approach in the Landau gauge: Infrared propagators in harmony with the lattice results}, Phys. Rev. D 78, 065047 (2008).

\bibitem{Vandersickel11}
N. Vandersickel, {\it A Study of the Gribov-Zwanziger action: from propagators to glueballs}, arXiv 1104.1315.

\bibitem{Brito24}
G. P. de Brito and A. D. Pereira, {\it Infrared gluon propagator in the refined Gribov-Zwanziger scenario at one-loop order in the Landau gauge}, Phys. Rev. D 110, 074005 (2024).

\bibitem{Brito25}
G. P. de Brito, P. De Fabritiis, and A. D. Pereira,{\it Quark propagator at one loop in the refined Gribov-Zwanziger framework}, Phys. Rev. D 112, 014014 (2025).

\bibitem{Barrios24b}
N. Barrios, M. Peláez, M. S. Guimaraes, B. W. Mintz, and L. F. Palhares, {\it Ghost-gluon vertex in the presence of the Gribov horizon: General kinematics},  Phys. Rev. D 109, 094039 (2024).

\bibitem{Brito23}
G. P. de Brito, P. De Fabritiis, and A. D. Pereira, {\it Refined Gribov-Zwanziger theory coupled to scalar fields in the Landau gauge}, Phys. Rev. D 107, 114006 (2023).

\bibitem{Mintz18}
B. W. Mintz, L. F. Palhares, S. P. Sorella, and A. D. Pereira, {\it Ghost-gluon vertex in the presence of the Gribov horizon}, Phys.Rev.D 97, 034020 (2018).

\bibitem{vanBaal92}
P. van Baal, {\it More (thoughts on) Gribov copies}, Nucl. Phys. B369, 259 (1992).

\bibitem{Pelaez21Rev}
M. Peláez, U. Reinosa, J. Serreau, M. Tissier, and N. Wschebor, {\it A window on infrared QCD with small expansion parameters}, Rep. Prog. Phys. 84, 124202 (2021).

\bibitem{Curci76}
G. Curci and R. Ferrari, {\it On a class of Lagrangian models for massive and massless Yang-mills fields}, Nuov Cim A 32, 151 (1976).

\bibitem{Tissier10}
M. Tissier and N. Wschebor, {\it Infrared propagators of Yang-Mills theory from perturbation theory}, Phys. Rev. D 82, 101701(R) (2010). 
		
\bibitem{Tissier11}
M. Tissier and N. Wschebor, {\it Infrared safe perturbative approach to Yang-Mills correlators}, Phys. Rev. D 84, 045018 (2011). 



\bibitem{Pelaez13}
M. Pelaez, M. Tissier, and N. Wschebor, {\it Three-point correlation functions in Yang-Mills theory}, Phys. Rev. D 88, 125003 (2013).

\bibitem{Barrios24a}
N. Barrios, P. De Fabritiis, and M. Peláez, {\it Four-gluon vertex from the Curci-Ferrari model at one-loop order}, Phys. Rev. D 109, L091502 (2024).

\bibitem{Pelaez14}
M. Pelaez, M. Tissier, and N. Wschebor, {\it Two-point correlation functions of QCD in the Landau gauge}, Phys. Rev. D 90, 065031 (2014).

\bibitem{Pelaez15}
M. Pelaez, M. Tissier, and N. Wschebor, {\it Quark-gluon vertex from the Landau gauge Curci-Ferrari model}, Phys. Rev. D 92, 045012 (2015).

\bibitem{Figueroa22}
F. Figueroa and M. Peláez, {\it One-loop unquenched three-gluon and ghost-gluon vertices in the Curci-Ferrari model}, Phys. Rev. D 105, 094005 (2022).

\bibitem{Gracey19}
J. A. Gracey, M. Peláez, U. Reinosa, and M. Tissier, {\it Two loop calculation of Yang-Mills propagators in the Curci-Ferrari model}, Phys. Rev. D 100, 034023 (2019). 

\bibitem{Barrios20}
N. Barrios, M. Peláez, U. Reinosa, and N. Wschebor, {\it The ghost-antighost-gluon vertex from the Curci-Ferrari model: Two-loop corrections}, Phys. Rev. D 102, 114016 (2020).

\bibitem{Barrios21}
N. Barrios, J. A. Gracey, M. Peláez, and U. Reinosa, {\it Two-loop corrections to the QCD propagators within the Curci-Ferrari model}, Phys. Rev D 104, 094019 (2021).

\bibitem{Barrios22}
N. Barrios, M. Peláez, and U. Reinosa, {\it Two-loop three-gluon vertex from the Curci-Ferrari model and its leading infrared behavior to all loop orders},  Phys. Rev. D 106, 114039 (2022).


\bibitem{Reinosa14}
U. Reinosa, J. Serreau, M. Tissier, and N. Wschebor, {\it Yang-Mills correlators at finite temperature: A perturbative perspective}, Phys. Rev. D 89 (2014) 105016.

\bibitem{Reinosa15a}
U. Reinosa, J. Serreau, M. Tissier, and N. Wschebor, {\it Deconfinement transition in SU(N) theories from perturbation theory}, Phys. Lett. B 742, 61 (2015).  

\bibitem{Reinosa15b}
U. Reinosa, J. Serreau, M. Tissier, and N. Wschebor, {\it Deconfinement transition in SU(2) Yang-Mills theory: A two-loop study},  Phys. Rev. D 91, 045035 (2015).

\bibitem{Reinosa15c}
U. Reinosa, J. Serreau, and M. Tissier, {\it Perturbative study of the QCD phase diagram for heavy quarks at nonzero chemical potential}, Phys. Rev. D 92, 025021 (2015).

\bibitem{Reinosa16}
U. Reinosa, J. Serreau, M. Tissier, and N. Wschebor, {\it Two-loop study of the deconfinement transition in Yang-Mills theories: SU(3) and beyond}, Phys. Rev. D 93, 105002 (2016).

\bibitem{Reinosa17b}
U. Reinosa, J. Serreau, M. Tissier and A. Tresmontant, {\it Yang-Mills correlators across the deconfinement phase transition}, Phys. Rev. D 95, 045014 (2017).

\bibitem{Pelaez17}
M. Peláez, U. Reinosa, J. Serreau, M. Tissier, and N. Wschebor, {\it Small parameters in infrared quantum chromodynamics},  Phys. Rev. D 96, 114011 (2017).

\bibitem{Pelaez21}
M. Peláez, U. Reinosa, J. Serreau, M. Tissier, and N. Wschebor, {\it Spontaneous chiral symmetry breaking in the massive Landau gauge: realistic running coupling},  Phys. Rev. D 103, 094035 (2021). 

\bibitem{Pelaez23}
M. Peláez, U. Reinosa, J. Serreau, and N. Wschebor, {\it Small parameters in infrared QCD: The pion decay constant}, Phys. Rev. D 107, 054025 (2023).

\bibitem{Pelaez25}
M. Peláez, U. Reinosa, J. Serreau, M. Tissier, and N. Wschebor, {\it Scenario for quark confinement from infrared safe Yang-Mills dynamics}, Phys.Rev.D 112, 094017 (2025).

\bibitem{DeFabritiis26}
D. Dudal, P. De Fabritiis, and S. Stouten, {\it Non-Abelian Casimir energy in the Curci-Ferrari model through a functional approach},  J. High Energ. Phys. 2026, 87 (2026).

\bibitem{Reinosa17}
U. Reinosa, J. Serreau, M. Tissier, and N. Wschebor, {\it How nonperturbative is the infrared regime of Landau gauge Yang-Mills correlators?}, Phys. Rev. D 96, 014005 (2017).

\bibitem{Oribe25}
S. Oribe, M. Peláez, and U. Reinosa
, {\it Infrared-safe Minkowskian Curci-Ferrari model}, Phys. Rev. D 112, 014005 (2025).


\bibitem{Maas19a}
A. Maas, {\it Quenched SU(2) adjoint scalar propagator in
minimal Landau gauge}, Phys. Rev. D 99, 054504 (2019).

\bibitem{Maas19b}
A. Maas, {\it Quenched SU(2) scalar-gluon vertex in minimal Landau gauge}, Phys. Rev. D 99, 114503 (2019).

\bibitem{Maas16}
A. Maas, {\it The quenched SU(2) fundamental scalar propagator in minimal Landau gauge}, Eur. Phys. J. C 76, 366 (2016).

\bibitem{Fradkin79}
E. Fradkin and S. H. Shenker, {\it Phase diagrams of lattice gauge theories with Higgs fields}, Phys. Rev. D 19, 3682 (1979).


\bibitem{Cucchieri:2008qm}
A.~Cucchieri, A.~Maas and T.~Mendes,
Phys. Rev. D \textbf{77}, 094510 (2008)
doi:10.1103/PhysRevD.77.094510
[arXiv:0803.1798 [hep-lat]].

\bibitem{Capri17}
M. A. L. Capri, D. Fiorentini, A. D. Pereira and S. P. Sorella,
{\it A non-perturbative study of matter field propagators in Euclidean Yang{\textendash}Mills theory in linear covariant, Curci{\textendash}Ferrari and maximal Abelian gauges},
Eur. Phys. J. C 77, 546 (2017).

\bibitem{Capri14}
M. A. L.~Capri, M. S. Guimaraes, I. F. Justo, L. F.  Palhares and S. P. Sorella,
{\it Properties of the Faddeev-Popov operator in the Landau gauge, matter confinement and soft BRST breaking}, 
Phys. Rev. D 90, 085010 (2014).





\end{thebibliography}
\end{document}